\documentclass[12pt, a4paper]{article}

\pdfoutput=1

\usepackage{amsmath}
\usepackage{amsfonts}
\usepackage{amssymb}
\usepackage{cite}
\usepackage{multicol,multirow}
\usepackage{graphicx, physics
			}
\graphicspath{{figs/}}
\usepackage{graphics, epsfig}

\usepackage{mcite}

\usepackage{stackengine}
\newcommand\textsub[1]{\stackengine{-.5ex}{}{\scriptsize#1}{O}{l}{F}{F}{L}}
\renewcommand\textsub[1]{\stackengine{-.4ex}{}{\tiny#1}{O}{l}{F}{F}{L}}

\def\DE{D\textsub{E}}
\def\DB{D\textsub{B}}
\def\DX{D\textsub{X}}

\usepackage[usenames,dvipsnames]{color}

\newcommand{\sss}[1]{{\scriptscriptstyle{#1}}}

\newcommand{\uPl}{\mathrm{Pl}}

\newcommand{\usssPl}{\sss{\uPl}}

\newcommand{\Mp}{M_\usssPl}

\newcommand{\beq}{\begin{equation}}
\newcommand{\eeq}{\end{equation}}
\newcommand{\bea}{\begin{eqnarray}}
\newcommand{\eea}{\end{eqnarray}}

\newlength{\wsingfig}
\newlength{\wdblefig}
\newlength{\wquadfig}
\newlength{\wtriplefig}
\usepackage{colortbl}
\definecolor{summersky}{cmyk}{0.71,0.33,0,0.14}
\definecolor{flamingo}{cmyk}{0,0.51,0.71,0.14}
\definecolor{rp}{cmyk}{0.2, 1, 0.6, 0}
\definecolor{pacificblue}{cmyk}{0.95,0.3,0, 0.19}
\definecolor{gray60}{cmyk}{0.4,0.4,0,0.8}

\numberwithin{equation}{section}

\usepackage[colorlinks,bookmarks=false, hypertexnames=true]{hyperref}
\hypersetup{
	breaklinks=true,
	pdfstartview={FitH},    
	colorlinks=true,       
	linkcolor=blue,          
	citecolor=red,        
	filecolor=magenta,      
	urlcolor=red,           
	anchorcolor=green,      
	linktocpage=true
}

\def\bfx{{\bf x}}

\DeclareMathAlphabet{\mathpzc}{OT1}{pzc}{m}{it}
\usepackage[normalem]{ulem}
\newcommand\redsout{\bgroup\markoverwith{\textcolor{red}{\rule[0.4ex]{2pt}{2pt}}}\ULon}

\begin{document}


\begin{center}

{\bf\large Primordial Helical  Magnetic Fields from Inflation?}
\\[0.5cm]

{ Alireza Talebian \footnote{talebian@ipm.ir}, 
Amin Nassiri-Rad \footnote{amin.nassiriraad@ipm.ir}, 
Hassan Firouzjahi \footnote{firouz@ipm.ir},
}
\\[0.5cm]
 
 {\small \textit{$^1$School of Astronomy, Institute for Research in Fundamental Sciences (IPM) \\ P.~O.~Box 19395-5531, Tehran, Iran
}}\\

\today
\end{center}

\vspace{.3cm}
\hrule
\begin{abstract}
We revisit the mechanism of helical magnetogenesis during inflation with a parity violating interaction using the formalism of stochastic inflation.  
One of the polarization of the gauge field undergoes tachyonic growth leading to the generation of helical magnetic fields. We obtain the Langevin equations associated with the electromagnetic fields which are in the form of Ornstein-Uhlenbeck stochastic differential equations. Consequently, the  tachyonic growth of the helical magnetic fields is balanced   by a mean-reverting process of stochastic dynamics such that the magnetic fields settle down to an equilibrium state with the amplitude  smaller than what is obtained in the absence of the stochastic noises. Working in the parameter space of the model where both the backreaction and the strong coupling problems are under control the model does not provide large enough seed to be amplified by the galactic dynamo as the source of  the magnetic fields observed on cosmological scales.
\end{abstract}

\vspace{0.3cm}
\hrule


\tableofcontents
\section{Introduction}
\label{sec:intro}

Magnetic fields are present throughout the observable Universe:  in stars, in the interstellar medium, in galaxies, and in clusters of galaxies. However, 
from astrophysical point of view, the origin of magnetic fields on large cosmological scales  is still mysterious. 
On all scales an initial magnetic seed with a sufficient strength is needed. Seed fields may be generated with different  strengths due to a variety of processes \cite{kulsrud1997protogalactic,Furlanetto:2001gx}. There has been a lot of debates whether seed fields can be produced by battery mechanism (charge separation processes, separation of charges and production of currents) during galaxy and cluster formation \cite{zeldovich1983magnetic,parker2019cosmical} or whether seed fields with primordial origin are needed. Both scenarios are currently under active considerations  \cite{Giovannini:2003yn, Giovannini:2017rbc, Subramanian:2015lua, Durrer:2013pga, Widrow:2002ud}. Several mechanisms have been proposed for the origin of primordial seed fields, ranging from cosmological phase transitions \cite{vachaspati1991magnetic} to the inflationary production of magnetic fields \cite{Turner:1987bw,Ratra:1991bn,VargasMoniz:2010upl,Ng:2014lyb,Gasperini:2017fqw,Gorbar:2021rlt,Gorbar:2021zlr,Bamba:2021wyx,Maity:2021qps,Giovannini:2021due,Giovannini:2021thf,Giovannini:2021dso,Giovannini:2021xbi,Giovannini:2020zjo}.
For a review of proposed scenarios we refer the reader to \cite{Grasso:2000wj,Giovannini:2003yn, Vachaspati:2020blt}.

We are interested in two classes of observations at two different scales that put  constraints on magnetogenesis scenarios.  The first is Galactic Magnetic Fields (GMF) with the amplitude of order of $\sim \mu{\rm G}$ and the second is the lower bound $10^{-16}{\rm G}$ on Inter Galactic Magnetic Fields (IGMF) on ${\rm Mpc}$ scales.  Not only the mechanism behind the 
generation of magnetic fields with large correlation length $L_{\rm B} \gtrsim 1 {\rm Mpc}$ remains mysterious \cite{wielebinski2005cosmic
,Durrer:2013pga}, but also on smaller scales, $L_{\rm B} \lesssim 1 {\rm Mpc}$, the origin of an initial magnetic seed with a sufficient strength (to be amplified by either astrophysical or primordial processes) is still unknown.

While the astrophysical origins for the observed magnetic fields are not excluded, the detection of magnetic fields with a correlation length $ \, \gtrsim \,{\rm Mpc}$ in cosmic voids ~\cite{Neronov:1900zz, Dolag:2010ni, Essey:2010nd, Taylor:2011bn,Caprini:2015gga}
has rekindled the interest in the construction of inflationary mechanisms of magnetogenesis.  Cosmic inflation allows us to imagine that quantum fluctuations of magnetic fields are stretched beyond the horizon which later seed the observed magnetic fields with very large coherent length \cite{Turner:1987bw,Ratra:1991bn}, an opportunity which is not available in models of the early universe without inflation. However, the conformal invariance of Maxwell theory implies that magnetic fields can not be generated in an expanding Universe \cite{Turner:1987bw,Ratra:1991bn, Parker:1968mv}. A simple way to generate the magnetic fields during inflation is to introduce an interaction between the inflaton and the electromagnetic fields which breaks the conformal invariance. Since the violation of gauge invariance generally gives rise to ghost-like instabilities\cite{Himmetoglu:2008zp,Himmetoglu:2009qi}, the mechanisms which preserve gauge invariance while breaking the conformal invariance have gained most of the attentions.

The best-studied model of inflationary magnetogenesis is the so-called Ratra model\footnote{For conciseness we refer to this model as the ``Ratra model''.} \cite{ Ratra:1991bn, Gasperini:1995dh, Bamba:2003av, Martin:2007ue, Demozzi:2009fu, Kanno:2009ei, Emami:2009vd, Fujita:2012rb, Giovannini:2013rme, Fujita:2013pgp, Ferreira:2013sqa, Ferreira:2014hma, Kobayashi:2014sga, BazrafshanMoghaddam:2017zgx,  Sobol:2020lec, Barnaby:2012tk} in which an interaction between the electromagnetic field and the inflaton (or a spectator) field  was introduced. The action contains the non-minimal coupling   $I^2(\phi)F_{\mu\nu}F^{\mu\nu}$ where $\phi$ is the inflaton field, $F_{\mu \nu}$ is the electromagnetic field strength and the conformal coupling $I(\phi)$ was added to break the conformal invariance. The elementary versions of the model suffers from two main problems: the strong coupling problem and  the backreaction problem \cite{Gasperini:1995dh,Demozzi:2009fu,Fujita:2012rb,
Martin:2007ue,Kanno:2009ei, 
Emami:2009vd}. Although for some ranges of the parameter space  both problems are bypassed but the generated magnetic field is not stronger than $10^{-32}$~G~\cite{Demozzi:2009fu} at $1$~Mpc today.  Another well-studied mechanism resides on the combination of a Ratra-like coupling with an axion-like coupling, \textit{i.e.} $L \supseteq I^2(\phi) \left(F^{\mu \nu}F_{\mu \nu}+\gamma F^{\mu \nu}\tilde{F}_{\mu \nu}\right)$ 
in which $\gamma$ is a constant. This Lagrangian generates helical magnetic fields. After inflation and before recombination, as the plasma is highly turbulent with a large Reynolds number, the inverse cascade process plays a significant role in the subsequent evolution of magnetic fields and their coherence length~\cite{Banerjee:2004df}.  Taking into account the constraints from non-Gaussianities \cite{Caprini:2017vnn} and induced gravitational waves \cite{Caprini:2014mja} in this model, the scenario can satisfy the observational lower bounds on IGMF, while providing a seed for the galactic dynamo to generate GMF. Caprini and Sorbo \cite{Caprini:2014mja} have claimed that the model  can provide a magnetic field amplitude of the order of $10^{-19} {\rm G}$ on the ${\rm Mpc}$ scale \cite{Caprini:2014mja}. This comes at the price of a low energy scale of inflation, ranging from $10^{5}$ to $10^{10}~ \rm GeV$~.

In our previous work \cite{Talebian:2020drj} we have studied the effects of electromagnetic noises on the generation of primordial magnetic fields in the Ratra-like model using the mechanism of stochastic inflation. It was shown that the stochastic effects can play important roles which affect the previous estimations on the amplitude of backreactions on the inflation dynamics, yielding large enough seeds required for magnetogenesis. 
Motivated by the non-trivial contributions of the stochastic noises on the system containing gauge fields, one may expect that the stochastic effects can play important roles in axion magnetogenesis setup as well. In this paper, we revisit the axion magnetogenesis  model taking into account the stochastic noises of the electromagnetic fields. We show that indeed the stochastic effects can significantly  modify  the previous results for helical magnetogenesis.

The rest of the paper is organized as follows. In Sec.~\ref{model}, the magnetogenesis mechanism in model of inflation with the parity violating interaction  is reviewed and the relevant results of previous works, \textit{e.g.} \cite{Caprini:2014mja,Caprini:2017vnn} are presented. In Sec.~\ref{sto} we revisit the setup taking into account stochastic noises and derive the Langevin equations of the electric and magnetic fields and investigate the parameters of the model. In Sec.~\ref{BToday} we discuss the observational constraints on the magnetic fields at the present time and search the parameter space of the model where the constraints are satisfied. Section~\ref{Conclusion} is devoted to the summary and conclusions while many technicalities associated with the stochastic noises and their correlations and the cosmological evolution of the magnetic fields are relegated to the appendices.

\section{The Model}
\label{model}

The setup is based on a  hybrid of Ratra and axion models.  The electromagnetic Lagrangian density ${\cal L}_{\rm EM}$
consists of a $U(1)$ gauge field $A_\mu$  coupled to an axionic inflaton field 
$\phi$ via,
\begin{equation}
\label{LEM}
{\cal L}_{\rm EM}=-\dfrac{1}{4}
I^2(\phi) 
\, 
\bigg(
 F_{\mu\nu}F^{\mu\nu} + 
 \frac{\gamma}{2} 
 F_{\mu\nu}\tilde{F}^{\mu\nu}
\bigg)
\,,
\end{equation}
where $F_{\mu\nu}=\partial_\mu A_\nu-\partial_\nu A_\mu$ is the field strength and $\tilde{F}^{\mu \nu} \equiv \frac{\epsilon^{\mu \nu \alpha \beta}}{2\sqrt{-g}}F_{\alpha \beta}$ is its dual with $\epsilon^{0123}=1$. Here, we choose the constant parameter $\gamma <0$ without loss of generality. Since the energy density of the  electromagnetic field is exponentially diluted during the 
quasi de Sitter inflationary expansion, the conformal coupling $I(\phi)$ is added to break the conformal invariance. With this conformal coupling the energy is continuously pumped from the inflaton sector to the gauge field sector so the electromagnetic energy density survives the exponential dilution.  Although the conformal coupling is a function of $\phi$, but the latter itself is a function of time so we consider the following phenomenological ansatz
for the conformal coupling
\begin{align}
\label{f}
I(\tau) = I_{\rm e}
\left(
\dfrac{\tau}{\tau_{\rm e}}
\right)^n \,,
\end{align}
where the conformal time $\tau$ is related to the cosmic time $t$ via the scale factor $a$ as $\dd t = a \, \dd \tau$,  $\tau_{\rm e}$ and $I_{\rm e}$ are the corresponding terminal values at the end of inflation. This coupling was employed
extensively in the context of anisotropic inflation \cite{Watanabe:2009ct,Watanabe:2010fh,Bartolo:2012sd,Emami:2010rm, Emami:2013bk, Emami:2014tpa, Emami:2015qjl, Abolhasani:2015cve}, and the generation of primordial magnetic field during inflation
\cite{ Ratra:1991bn
, Gasperini:1995dh, Bamba:2003av, Martin:2007ue, Demozzi:2009fu, Kanno:2009ei, Emami:2009vd, Barnaby:2012tk, Fujita:2012rb, 
Giovannini:2013rme,Fujita:2013pgp,Ferreira:2013sqa,Ferreira:2014hma,Kobayashi:2014sga, BazrafshanMoghaddam:2017zgx,  Sobol:2020lec}.
The spectral index of the magnetic field is controlled by the the parameter $n$, so that a scale invariant magnetic field can be obtained for the cases  $n=3$ or $n=-2$.

To study the magnetogenesis in the presence of Lagrangian ${\cal L}_{\rm EM}$ \eqref{LEM}, we start by the following action:
\begin{eqnarray}
\mathcal{S} = \int \mathrm{d}^4 x \, \sqrt{-g} \, \bigg[ \dfrac{\Mp^2}{2} R 
- \dfrac{1}{2} g^{\mu \nu} \partial_{\mu} \phi  \partial_{\nu} \phi - V(\phi) 
+ {\cal L}_{\rm EM} \bigg] \,,
\label{action}
\end{eqnarray}
in which $R$ is the Ricci scalar and $\Mp \equiv (8\pi G)^{-\frac{1}{2}}$ is the reduced Planck mass with $G$ being the Newton constant.

We assume that the electromagnetic fields are purely excited quantum mechanically. This means that the electromagnetic fields have no background components so they do not contribute to the background energy density. The background is given by a spatially flat, Friedmann-Lemaitre-Robertson-Walker (FLRW) universe, described by the line-element
\begin{equation}
\label{metric}
\dd s^2 = a^2(\tau) \big(-\dd \tau^2 +  \ \dd \bfx \cdot \dd \bfx \big) \,,
\end{equation}
in which the relation $a \simeq -\left( H \tau \right)^{-1}$ can be used with good accuracy where $H$ is the Hubble expansion rate during inflation. 

It is more convenient to work in the temporal-Coulomb gauge  
$A_0=\partial_i A_i=0$ and define the electric and magnetic fields as
\begin{align}
	\label{electric}
	E_i \equiv-I\frac{\partial_\tau{A_i}}{a^2} \,,
	\hspace{1cm}
	B_i \equiv I\frac{\epsilon_{ijk}\partial_j{A_k}}{a^2} \, .
\end{align}
With the above definitions, the  Friedmann and Klein-Gordon (KG) equations  take the  following form
\begin{align}
\label{Friedmann}
&3 \Mp^2 H^2 = \dfrac{1}{2}\dot{\phi}^2+V +\rho_{\rm EM} \,;
\hspace{2.5cm} \rho_{\rm EM} \equiv \dfrac{1}{2}\left(E^2+B^2\right) \,,
\\
\label{KG}
&\ddot{\phi}+3\,H\,\dot{\phi}-\dfrac{\nabla^{2}}{a^2}\phi+V_{,\phi}= S_{\rm EM} \,;
\hspace{1.8cm}
S_{\rm EM}\equiv \dfrac{1}{\dot{\phi}a}\dfrac{I'}{I}\left(
E^2 - B^2 -2\gamma\boldsymbol{E} \cdot \boldsymbol{B}
\right)
\,,
\end{align}
where a dot (prime)  denotes the derivative with respect to the cosmic time 
$t$ (conformal time $\tau$).  Note that $\rho_{\rm EM}$ represents the electromagnetic field energy density while $S_{\rm EM}$ is the backreaction source of the electromagnetic fields on the KG field equation.

Finally, the Maxwell equations with the Bianchi identities read as
\begin{align}
\label{B-eq}
&\dot{\bf B}+H(2+n){\bf B}=-\frac{{\boldsymbol \nabla} \times {\bf E}}{a} \,,
\\
\label{E_eq}
&\dot{\bf E}+\gamma \dot{\bf B}+ H(2-n)\left(
{\bf E}+\gamma {\bf B}
\right) =-\frac{{\boldsymbol \nabla} \times \left({\bf B}+\gamma {\bf E}\right)}{a}
\,.
\end{align}

There are two important issues which should be taken into account when constructing a scenario of magnetogenesis during inflation: the strong coupling problem and the electric field backreaction problem. Looking at the electromagnetic action we realize that the gauge coupling is $I(\phi)^{-1}$ so in order for the perturbative field theory to be trusted we require that $I(\phi) \ge 1$ for all time during inflation. With the phenomenological ansatz given in Eq. (\ref{f}), we need $n>0$ in order to avoid the strong coupling problem. Furthermore, we set 
$I_{\rm e} = 1$ in Eq. (\ref{f})  such that we recover the standard Maxwell theory 
 at the end of inflation when the inflaton decays through the (p)reheating process. 
The backreaction problem, on the other hand, is associated to the fact that for some regions of parameter space the electric field is enhanced so efficiently that its energy density can dominate over the inflaton potential, terminating inflation prematurely \cite{Demozzi:2009fu}. As studied in  \cite{Talebian:2020drj}, to avoid the backreaction and the strong coupling problems we require $\frac{1}{2} < n < 2$ which will be considered in this work as well. As studied in  \cite{Talebian:2020drj} (see also \cite{Fujita:2017lfu, Talebian:2019opf}), stochastic effects have important implications for these two problems. 
The stochastic  noises cause the solutions of electromagnetic fields to settle down to an equilibrium state in such a way that (for an acceptable range of parameter space) not only the backreaction effects can be under control but also an acceptable amount of magnetic field is generated.

To have  the backreaction effects under control, we assume that the gauge field contributions do not destroy the dynamics of the
inflaton field and the background geometry given in  equations \eqref{Friedmann} and \eqref{KG} respectively. This means
\begin{align}
\label{constraint1}
\Omega_{\rm EM} &\equiv \dfrac{\rho_{\rm EM}}{3\Mp^2H^2} \ll 1 \,,
\\
\label{constraint2}
R_{S} &\equiv \abs{\dfrac{S_{\rm EM}}{3H\dot{\phi}}}  \ll 1 \, .
\end{align}
Both of the above conditions must be satisfied during inflation.  We check these conditions in the context of stochastic formalism and specify the 
the allowed regions of the parameter space. 

To study the behaviour of electromagnetic fields in this model, we look at the quantum fluctuations of the gauge field during inflation. Defining the  
canonical field as $\tilde{A}_i \equiv I A_i$ and going to the Fourier space 
we expand $\tilde{A}_i$ in terms of the creation and annihilation operators $a_{\bf k}$ and $a^{\dagger}_{\bf k}$ 
as follows 
\begin{equation}
\label{A_tilde}
\tilde{A}_i=\sum_{\lambda=\pm} \int \dfrac{\dd ^3k}{(2\pi)^3}e_i^\lambda({\bf k}) \bigg(
v_{k,\lambda}(\tau) \, a_{{\bf k},\lambda} + v^*_{k,\lambda}(\tau) \,  a^\dagger_{-{\bf k},\lambda}
\bigg) e^{i{\bf k}.{\bf x}}
\,,
\end{equation}
in which $v_{k,\lambda}$ is the mode function and  
$\boldsymbol{e}^\lambda$ are the circular polarization vectors satisfying the relations
\begin{eqnarray}
\label{Orthogonality}
\boldsymbol{e}_\lambda(\hat{\boldsymbol{k}}).\boldsymbol{e}_{\lambda'}^*(\hat{\boldsymbol{k}})&=&\delta_{\lambda\lambda'} \,,
\nonumber\\
\boldsymbol{\hat{k}}.\boldsymbol{e}^\lambda(\hat{\boldsymbol{k}}) &=& 0 \,,
\nonumber\\
i\hat{\boldsymbol{k}} \times \boldsymbol{e}^\lambda &=& \lambda \boldsymbol{e}^\lambda \,,
\label{k-cross-e}
\nonumber\\
\boldsymbol{e}_\lambda(\hat{\boldsymbol{k}}) &=& \boldsymbol{e}^*_{\lambda}(-\hat{\boldsymbol{k}}) \,,
\nonumber\\
\sum_{\lambda = \pm} e_i^{\lambda}(\hat{\boldsymbol{k}})~e_j^{\lambda *}(\hat{\boldsymbol{k}}) &=& \delta_{ij}-\hat{k}_i \hat{k}_j \,.
\end{eqnarray}
Substituting Eq. \eqref{A_tilde} in the action and using the ansatz \eqref{f}, 
the equation for the  mode function is given by
\begin{align}
\label{v}
v_{k,\lambda}''+ \bigg(
k^2 + 2\lambda \xi \dfrac{k}{\tau}-\dfrac{n(n-1)}{\tau^2}
\bigg) v_{k,\lambda} = 0 \,;
\hspace{1cm}
\xi \equiv -n \, \gamma \, .
\end{align}
Here we have defined the instability parameter $\xi >0$ (remember that we have chosen $\gamma <0$ in the Lagrangian \eqref{LEM}). As we shall see $\xi$ is one key parameter of the model which controls the strength of tachyonic instability for the gauge field perturbations.  In the previous works of magnetogenesis based on the 
setup (\ref{LEM}) \cite{Anber:2006xt,Caprini:2014mja}, $\xi$ has taken to be in the range  
$\xi \sim {\cal O}(10)$.

The mode function $v_{k,\lambda}$ satisfying Eq. \eqref{v} evolves in three 
stages as follows.   During early times, $\tau \rightarrow -\infty$, the ultraviolet term $k^2$ dominates and the gauge quanta are in their Bunch-Davies vacuum. Later on, before horizon crossing, the term proportional to $\xi$ becomes important for the sub-horizon modes with $\abs{k\tau}  \lesssim \xi$.
Since $\tau<0$ during inflation, then the mode function with positive helicity 
is exponentially amplified whereas the mode of opposite helicity does not experience  such an amplification. In order for an efficient enhancement to take place one requires $\xi \gg 1$  in which a net chirality in the gauge field perturbations  is generated. Finally, as
$\tau \rightarrow 0$ the last term in the parenthesis takes over as in conventional models of inflation based on scalar field dynamics. 

The above three-stage  processes can be addressed by a function with three arguments such as the Whittaker functions. Actually, the solutions to this equation are given by a linear combination of Whittaker functions $W_{\mu,\nu}(z)$ and $M_{\mu,\nu}(z)$ with the coefficients determined by the initial conditions. Imposing the standard Bunch-Davies solutions at early times\footnote{$W_{\mu,\nu}(z) \rightarrow z^\mu e^{-z/2}$ for $z \rightarrow \infty$.} 
$-k\tau \rightarrow \infty$, the solution of \eqref{v} is given by
\begin{align}
\label{v-mode}
v_{k,\lambda}(\tau) = \dfrac{e^{\frac{{\lambda \pi \xi}}{2}}}{\sqrt{2k}} \, W_{\mu,\nu}(2ik\tau) \,; \hspace{1cm}
\mu \equiv -i\lambda \xi \,,
\hspace{1cm} \nu \equiv n-1/2 \,.
\end{align}
For $\gamma=0$ it is easy to check that the above mode function coincides with the  well-known mode function in terms of the Hankel functions\footnote{$W_{0,\nu}(z)=\frac{\sqrt{\pi z}}{2}i^{\nu+1}H^{(1)}_\nu(\frac{iz}{2})$ in which $H^{(1)}(x)$ is the Hankel function of the first kind.} used in 
earlier studies such as in  \cite{Talebian:2020drj}.

During the second stage in which the second term in the parenthesis in Eq. \eqref{v} dominates, the sub-horizon modes  with $\lambda=+$ are amplified. 
In the regime $\left|k\,\tau\right|\ll\xi$ the solution \eqref{v-mode}  is approximated to  \cite{Durrer:2010mq,Caprini:2014mja}
\begin{equation}\label{sol:a:full}
v_{k}^{+}\left(\tau\right)\simeq\sqrt{-\frac{2\,\tau}{\pi}}e^{\pi\,\xi}K_{2\nu}\left(\sqrt{8\,\xi\, \abs{k\, \tau}}\right)\,,
\hspace{2cm}
\abs{k\tau} \ll \xi \,,~ \xi \gg 1 \,,
\end{equation}
where $K_{\nu}$ is the modified Bessel function of the second kind.
Subsequently, for $|k\,\tau| \ll 1/\xi$ we obtain
\begin{equation}\label{sol:a:approx}
v_{k}^{+}\left(k,\tau\right)\simeq\sqrt{-\frac{\tau}{2\,\pi}}\,e^{\pi\,\xi}\,\Gamma\left(2n-1\right)\left|2\,\xi\, k\, \tau\right|^{-\left(n-1/2\right)}\,,
\hspace{2cm}
\abs{k\tau} \ll 1/\xi \ll 1 \,.
\end{equation}
As seen, the  amplitude of the gauge field is exponentially enhanced via 
the instability parameter $\xi$.

Our first task is to calculate the amplitude of the generated magnetic field, its correlation length and the spectral index at the time of end of inflation $\tau=\tau_{\rm e}$.  In App.~\ref{B_Evolution} we have defined these quantities 
in Eqs. \eqref{B_intensity}, \eqref{L} and \eqref{n_B}, denoted respectively by
$B(\tau_{\rm e})$,   $L(\tau_{\rm e})$ and ${n_B}$.
Assuming  an instantaneous reheating scenario after inflation  
and denoting the values of $B(\tau_{\rm e})$  and $L(\tau_{\rm e})$ in the absence of stochastic effects  by $\bar{B}_{\rm {rh}}$ and $\bar{L}_{\rm {rh}}$,  the intensity of the magnetic field   is found to be\footnote{We estimate the reduced Planck mass in unit of Gauss as $\Mp^2 \simeq 3 \times 10^{56} \, {\rm G}$.}~\cite{Caprini:2014mja}
\begin{align}
\label{brh}
\bar{B}_{\rm rh} \simeq 1.9 \times 10^{53} \, {\rm G} \left(\dfrac{H}{\Mp}
\right)^2\,e^{\pi\,\xi}\, {\xi^{-5/2}} \,  {\sqrt{\Gamma(4+2\,n)\,\Gamma(6-2\,n)}} \,, 
\end{align}
while the  correlation scale is given by~\cite{Caprini:2014mja}
\begin{align}
\label{lrh}
\bar{L}_{\rm rh} 
 \simeq \frac{18\,\pi}{(3+2\,n)\,(5-2\,n)}\,\frac{\xi}{H}\,,
\end{align}
and the magnetic spectral index on large scales reads as
\begin{equation}
\label{nB1}
n_B=\frac{5}{2}-\Big| n-\frac{1}{2}\Big |\,.
\end{equation} 
Therefore the cases $n=3$ and $n=-2$ lead magnetic fields with scale invariant spectra. As mentioned before, to keep the backreaction and the strong coupling problems under control we require $\frac{1}{2} < n < 2$ so Eq. (\ref{nB1}) simplifies to
\begin{equation}
\label{nBb}
n_B= 3- n \, .
\end{equation}

For large enough values of $\xi$, the gauge field perturbations can induce sizeable gravitational waves \cite{Caprini:2014mja} and non-Gaussianities \cite{Caprini:2017vnn} which are under observational constraints on CMB scales  \cite{Akrami:2018odb}. Therefore, the tensor-to-scalar ratio $r_{\rm t}$ and the equilateral configuration non-Gaussianity $f_{\rm NL}^{\rm equil}$ can be used to express the Hubble parameter $H$ in terms of the model parameter $n$ and $\xi$. According to \cite{Caprini:2014mja} and \cite{Caprini:2017vnn}, we have
\begin{align}
\label{H_Caprini}
\dfrac{H}{\Mp} &\simeq e^{-\pi \xi}\Big(
\dfrac{r_{\rm t}{\cal P}_\zeta}{p^t(n)}
\Big)^{1/4}\xi^{3/2}
\\
\label{H_Caprini_2017}
\dfrac{H}{\Mp} &\simeq e^{-\pi \xi}\Big(
\dfrac{f_{\rm NL}^{\rm equil} \, {\cal P}_\zeta^2}{p^f(n)}
\Big)^{1/6}\xi^{3/2} \, ,
\end{align}
where ${\cal P}_\zeta \simeq 2.1 \times 10^{-9}$ is the amplitude of the scalar perturbations and the functions  $p^t(n)$ and $p^f(n)$ are defined in \cite{Caprini:2014mja} and \cite{Caprini:2017vnn}, respectively. The above relations are obtained  for $\xi \sim {\cal O}(10)$ which leads to a very small energy scale of inflation. However, in the following analysis and in the presence of stochastic noises,  
we show that for $\xi\sim{\cal O}(10)$ there are significant backreactions on Klein-Gordon equation  which spoil the  inflationary dynamics. To bypass this issue, the upper bound $\xi \lesssim 3$ must be considered which is consistent with the findings of \cite{Durrer:2010mq} and  \cite{Salehian:2020asa}. We confirm that for $\xi \lesssim 3$ the usual vacuum tensor perturbations has the dominant contribution in $r_{\rm t}$ so that the energy scale of inflation can take higher values in contrast 
to the conclusion  of  \cite{Caprini:2017vnn}.  

The above was a brief review of inflationary magnetogenesis in the setup with the action (\ref{action}) in conventional approach and in the absence of stochastic effects. In the next Section we revisit these conclusions in the context of stochastic inflation. 

\section{Stochastic analysis}
\label{sto}

In this section, we study the magnetogenesis mechanism  taking into account the  effects of stochastic noises.  We employ the formalism of stochastic inflation which is an effective theory for the long wavelength 
modes \cite{Starobinsky:1994bd, Fujita:2014tja, Vennin:2015hra}.
In this formalism, the quantum fields are decomposed into the long and short wavelength modes. The long modes are the coarse grained perturbations on super-Hubble scales while 
the  short modes act as the stochastic forces for the evolution of the long modes at the time when they  leave the Hubble horizon. 
For light scalar perturbations  the amplitude of these stochastic noises is 
 $H/2\pi$  while  for the electromagnetic perturbations they show more non-trivial properties \cite{Talebian:2019opf,Talebian:2020drj}.


To perform stochastic analysis, we decompose the electric and magnetic fields into the long and short modes \cite{Sasaki:1987gy, Nambu:1988je, Nakao:1988yi}. Denoting  these fields collectively as $X=E_i,B_i$, we  write 
\begin{equation}
\label{X}
X=X_{\rm l}+\sqrt{\hbar} \ X_{\rm s} \, ,
\end{equation}
where the IR ($X_{\rm l}$) and UV ($X_s$) parts are decomposed via the step function $\Theta$ as 
\begin{equation}
\label{Xsl}
X_{\rm s,l}({\bf x},t)=\int \frac{{\rm d}^3k}{(2\pi)^3} \ \Theta \Big(
\pm k \, \mp \, k_{\rm c}
\Big) 
\ e^{i {\bf k}.{\bf x}} \ \hat{X}_k(t) \,,
\end{equation}
where the upper (lower) sign in \eqref{Xsl} corresponds to the short (long) modes and $k_{\rm c} \equiv \varepsilon \, a(t) H$ with $\varepsilon \ll 1$ being a small cutoff parameter. In addition  $\hat{X}_{\bf k}$ is the quantum operator expanded as
\begin{equation}
\hat{X}_{\bf k}=a_{\bf k} X_k+a^\dagger_{-\bf k} X_{-k}
\,;
\hspace{2cm}
\left[
a_{{\bf k}',\lambda'},a^\dagger_{{\bf k},\lambda}
\right] = (2\pi)^3\delta_{\lambda \lambda'} \, \delta^3({\bf k}-{\bf k}') \,,
\end{equation}
where  $a_{\bf k}$ and $a^{\dagger}_{\bf k}$ are the usual ladder operators and $X_k$ is the Fourier component of the fields. Note that these ladder operators are the same for electric, magnetic and the gauge fields.

To perform stochastic calculus, it is more convenient to use the dimensionless variables $\cal X = \cal B, \cal E$ associated to the long mode perturbations of the  electric and magnetic fields defined via 
\begin{align}
\label{X_dimless}
{\cal X} \equiv \dfrac{X_{\rm l}}{H \Mp}
\,
\,.
\end{align}
Substituting Eq. \eqref{X} into equations \eqref{B-eq} and \eqref{E_eq} and expanding for the long modes, $i.e.$ $k < k_{\rm c}$, we find the Langevin equations for the electric and magnetic fields. 
More specifically, neglecting the terms proportional to the gradients of the fields or the slow-roll parameters, we obtain
\begin{align}
\label{langevinB}
{\cal B}'_i &= -(2+n){\cal B}_i+ \hat{\sigma}^{_B}_i(N) \,,
\\
\label{langevinE}
{\cal E}'_i &= -(2-n){\cal E}_i+2 n\gamma {\cal B}_i+\hat{\sigma}^{_E}_i(N) \,,
\hspace{1cm} i = 1,2,3 \ ,
\end{align}
where the index $i$ represents the spatial components of the fields and the prime here and below denotes  the derivative with respect to the $e$-folding number, ${\rm d}N=H {}\rm dt$. The quantum noises $\hat{\sigma}^{_X}(N)$, emerging from the UV modes, are defined as
\begin{equation}
\label{sigmaX}
\hat{\sigma}^{_X}({\bf x},t) = -\dfrac{{\dd}k_{\rm c}}{\dd t}\int \frac{{\rm d}^3k}{(2\pi)^3} \, \delta( k-k_{\rm c}) \, e^{i {\bf k}.{\bf x}} \, \hat{\cal X}_k(t) \,.
\end{equation}

Both the electric and magnetic noises are determined via the mode function of the gauge field \eqref{v-mode}. We
are ultimately interested in the superhorizon behaviour of the above mode function which controls the behaviour of quantum noises $\hat{\sigma}^{_X}$. The properties of the electromagnetic noises and their correlations are studied in  In App.~\ref{noise}.  Here we rewrite \eqref{X_correlation} in terms of the number of $e$-fold as 
\begin{align}
\left \langle
\hat{\sigma}^{_X}_i(N_1) \, \hat{\sigma}^{_X}_j(N_2)
\right \rangle
&=
D\textsub{X}^2 \, \delta_{ij} \, \delta(N_1-N_2) \,,
\end{align}
where $D_{_X}$ is the diffusion coefficient defined by
\begin{align}
\label{DX}
D\textsub{X}^2 &\equiv  \dfrac{1}{18\pi^2} \dfrac{{\rm d}k_{\rm c}^3}{{\rm d}N} \ \sum_{\lambda}^{}\big|X_\lambda(N,k_c)\big|^2 \, .
\end{align} 
By using the small argument behaviour of the mode function \eqref{v-mode}\footnote{$W_{\mu,\nu}(z) \rightarrow z^{1/2-\nu} \Gamma(2\nu)/\Gamma(\nu-\mu+1/2)$ for $z \rightarrow 0$.}, the diffusion coefficients have been calculated in \eqref{DXX} which for $\xi \gg1$ yields\footnote{We have used the approximation $\abs{\Gamma(x+iy)}^2 \simeq \pi/(y \sinh(\pi y))$ for $y \gg x$.}
\begin{align}
\label{DB}
\DB
&\simeq  \dfrac{e^{\pi\xi} \sqrt{\xi}}{\pi \sqrt{3\pi} }   \dfrac{\Gamma(2n-1)}{2^{n}} \ \dfrac{H}{\Mp} \varepsilon^{n_B}
\\
\label{DE}
\DE
&= \DB \dfrac{(2n-1)}{\varepsilon} \, ,
\end{align}
where the magnetic field spectral index is defined in 
Eq. (\ref{nBb}), $n_B= 3-n$.  

There are some important comments which we discuss here. First, the amplitude of the diffusion coefficient of the electric field  is stronger than that of the magnetic field by a factor $\varepsilon^{-1}$. This is the reason why it is always the backreaction from the electric field which spoils the slow-roll inflation. Second, while the parameter $\varepsilon$ is employed in our analysis as a bookkeeping parameter to separate the long and short modes, but it appears on the diffusion coefficients as well. Curiously, the dependence on $\varepsilon$ for both diffusion coefficients are exactly determined by the scale dependency  of each perturbations, i.e. $D_B (D_E)$ is independent of $\varepsilon$ when $n=3 ~ (n=2)$. As we mentioned in previous section, it is a well-known result that the magnetic (electric) fields are scale invariant for  $n= 3 ~ (n=2)$.

Since the quantum  nature of these noises disappear for $\varepsilon \ll 1$ (see App.~\ref{quantum_nature} for more
details), one can express the quantum noises ${\hat{\sigma}}^{_X}(N)$ in terms of the classical normalized white noise $\sigma(N)$ as
\begin{align}
\label{sigmaXX}
{\hat{\sigma}}^{_X}_i(N) \equiv \DX \,\, \sigma_i(N) \,,
\end{align}
where
\begin{equation}
\langle \sigma_i(N) \rangle = 0 \,,
\hspace{3cm}
\langle \sigma_i(N_1) \, \sigma_j(N_2) \rangle = \delta_{ij} \, \delta(N_1-N_2)\,.
\end{equation}

Now we define a three dimensional (3D) Wiener (or Brownian) process with the components ${\rm W}_i$ associated with the noise $\sigma_i$ via
\begin{align}
\label{Wiener}
\dd {\rm W}_i(N) \, \equiv \, {\sigma}_i(N) \, \dd N \,,
\end{align}
and rewrite the stochastic differential  equations \eqref{langevinB} and \eqref{langevinE} in the following form
\begin{align}
\label{langvinB2}
\dd {\cal B}_i &= -(2+n){\cal B}_i \, \dd N + D\textsub{B} \, \dd {\rm W}_i(N) \,,
\\
\label{langvinE2}
\dd {\cal E}_i &= \Big(
-(2-n){\cal E}_i+2 n\gamma {\cal B}_i
\Big)\dd N+D\textsub{E} \, \dd {\rm W}_i(N) \,.
\end{align}

The second terms in each of the the above equations represent the effect of the random noises while  the first terms, proportional to $dN$, represent the classical drift term. For $1/2<n<2$, both of the above classical drift terms are  negative and the system is in the form of 
Ornstein-Uhlenbeck (OU) stochastic differential equations. The main feature of OU process is that the frictional drift force can be balanced by the random force so the stochastic fields ${\cal B}$ and ${\cal E}$ admit equilibrium states with long-term means  and  bounded variances (mean-reverting process). To be more precise, an OU process is a stationary Gauss-Markov process in which there is the tendency for the system to drift toward the mean value, with a greater attraction when the process is further away from the mean. For this process, the explicit dependence of the mean to the initial conditions is washed out over time and the system can be fully described by the drift and the diffusion coefficients.

To solve the  coupled Langevin equations (\ref{langvinB2}) and (\ref{langvinE2}), we go to a  basis that the equations are decoupled (see App.~\ref{diagonal} for more details). We have assumed that the electromagnetic fields are purely excited quantum mechanically. This means that the electromagnetic fields have no background components so we set the initial conditions for the electromagnetic fields to zero in Eqs.~\eqref{Bsol} and \eqref{Esol}, obtaining
\begin{align}
\label{Bsol2}
{\cal B}_i(N) &= 
\DB \, 
\int_{0}^{N} e^{(n+2) (N'-N)} \ 
\sigma_i(N') \, \dd N'
\\
\label{Esol2}
{\cal E}_i(N) &= 
-\gamma  {\cal B}_i(N)+\left(
\DE +\gamma\DB
\right)
\int_{0}^{N} e^{-(n-2) (N'-N)} \ \sigma_i(N') \, \dd N' \, .
\end{align}

Our main goal is to calculate various of the electric and magnetic correlation function (stochastic averages).  Using the following properties of the stochastic integrals \cite{evans2013introduction}
\begin{eqnarray}
\Big<
\int_0^{N'} G(N) \, \dd W(N) \int_0^{N'} F(N) \, \dd W(N')
\Big> 
&=& 
\int_0^{N'} G(N)F(N) \, \dd N \,,
\\
\Big<
\int_0^{N'} G(N) \, \dd W(N)
\Big> 
&=& 0 \,,
\end{eqnarray}
we can calculate the mean and the variance associated to ${\cal B}_i(N)$ and ${\cal E}_i(N)$. More specifically,
\begin{align}
\left< {\cal B}_i \right> &= \left< {\cal E}_i \right> =0\,,
\\
\label{averageB}
\left<{\cal B}_i^2(N)\right> &=
\frac{\DB^2}{2(n+2)}(1-e^{-2(n+2) N})
\\
\label{averageE}
\left<{\cal E}_i^2(N)\right>&=
\gamma^2\big<{\cal B}_i^2(N)\big>
-\frac{\gamma}{2}\DB(\DE +\gamma\DB)(1-e^{-4N}) 
\nonumber
\\
&\hspace{2cm}
+(\DE +\gamma\DB)^2  \times
\begin{cases}
\frac{1-e^{-2(2-n) N}}{2(2-n)} \,, \qquad & n \neq 2
\\
\\
N \,,\qquad & n = 2
\end{cases}
~\,,
\end{align}

\subsection{Backreaction condition}

Having calculated $\left<{\cal B}_i^2(N)\right> $ we can go ahead to look for 
the predictions of the model for the primordial magnetogenesis. However, before that, we should check the backreaction conditions induced on the background dynamics, parameterized by  Eqs. \eqref{constraint1} and \eqref{constraint2}. For the case $\xi <1$ there is no tachyonic instability from the parity violating term and, as shown in \cite{Talebian:2020drj},
the system is under control for the range  $\frac{1}{2}< n \leq 2$. However, in the current setup with $\xi>1$,
new backreactions from the tachyonic enhancement of the $+$ mode of the gauge field can be induced. As we demonstrate below, keeping the backreactions under control, one actually requires $\xi \lesssim3$.

Using the definition 
of ${\cal E}$ and ${\cal B}$ in  \eqref{X_dimless}, the backreaction constraints \eqref{constraint1} and \eqref{constraint2} are given by
\begin{align}
\label{Constrant1}
\Omega_{\rm EM} &
= \dfrac{1}{6}\left(
{\cal E}^2+{\cal B}^2 
\right) \ll 1 \,,
\\
\label{Constrant2}
R_{S} &
=\dfrac{n}{6\epsilon_\phi} \left(
{\cal E}^2-{\cal B}^2-2\gamma{\cal E}\cdot{\cal B}
\right) \ll 1 \,,
\end{align}
where the magnitude of the fields are given by ${\cal X} \equiv \left( \sum_{i=1}^{3}{\cal X}_i^2 \right)^{1/2}$ for $\cal X=\cal B, \cal E$ and the inflaton slow-roll parameter
$\epsilon_\phi$ is defined as
\begin{align}
\epsilon_\phi \equiv \dfrac{\dot{\phi}^2}{2\Mp^2H^2} \,.
\end{align}
Note that in general $\epsilon_\phi$ differs from the Hubble slow-roll parameter $\epsilon_H \equiv -\dot{H}/H^2$. Combining Eqs. \eqref{Friedmann} and \eqref{KG}, we find that
\begin{align}
\epsilon_H &=\epsilon_\phi+2\Omega_{\rm EM} -\dfrac{2n\gamma }{3}{\cal E}\cdot{\cal B} \, .
\end{align}
As seen, these two slow-roll parameters do not coincide in general, specially when the backreaction effects are significant. 

\begin{figure}[t]
	\centering
	\includegraphics[width=0.7\linewidth]{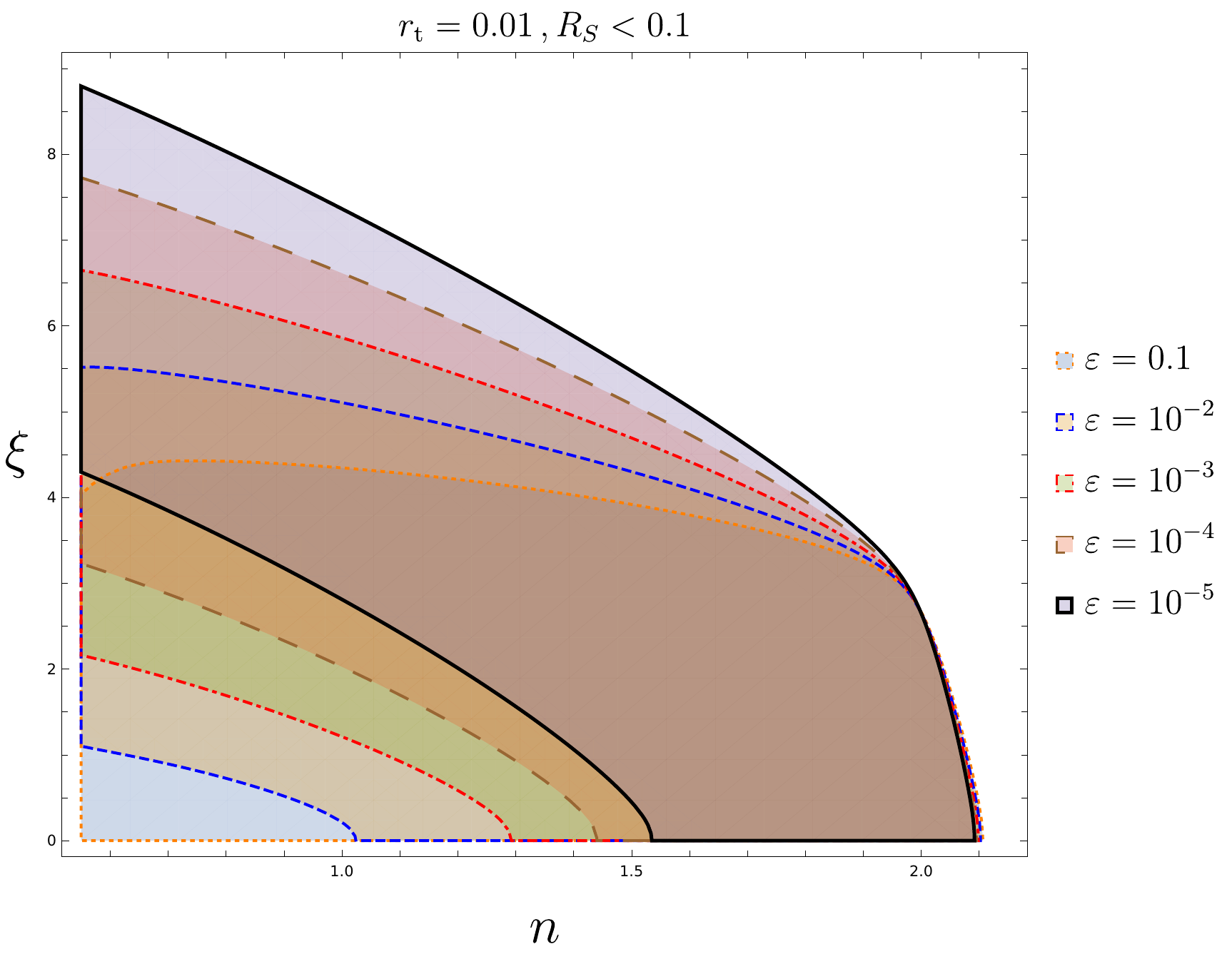}
	\caption{\footnotesize The allowed parameter space of $\xi-n$ where the backreaction effects are not significant, \textit{i.e} $R_S < 0.1$. To be conservative, for all values of $\varepsilon$, the backreactions can be neglected in the interval $\frac{1}{2}<n < 2$ when $\xi \lesssim 3$.}
	\label{fig:R-S}
\end{figure}

Now to estimate the backreaction effects, we note that  the constraint \eqref{Constrant2} is stronger than \eqref{Constrant1} by a slow-roll factor $\Omega_{\rm EM} \simeq \epsilon_\phi R_S$. This means that the backreactions from the electromagnetic field affect the dynamics of the inflaton field sooner than the background expansion rate.  Therefore the fractional energy density of the electromagnetic fields is subdominant and the constraint $R_S \ll 1$ must be checked first. For a fixed cutoff parameter $\varepsilon$, this constraint leads to a limited parameter space for $n$ and $\xi$. In Fig.~\ref{fig:R-S} we have plotted the allowed regions where the condition $R_S \ll 1$ is satisfied 
in the parameter space $\xi-n$. It is found that for the entire range $1/2<n<2$ we require $\xi \lesssim 3$ in order for the backreaction $R_S \ll 1$
to be satisfied.  
 Note that the constraint $\xi \lesssim 3$ is  obtained  in \cite{Durrer:2010mq,Salehian:2020asa} as well. In this range of parameter space, we can safely consider $\epsilon_H \simeq \epsilon_\phi$ to a very good accuracy. 

Now let us consider a slightly different setup, as studied  in \cite{Caprini:2017vnn},  in which a spectator field $\sigma$ other than the inflaton field is coupled to $F \tilde F$ term and $I = I(\sigma)$. Then one can parametrize the backreaction of gauge field on KG equation of the test field as
\begin{align}
\label{Constrant3}
R_{S}^\sigma &
\equiv
\dfrac{n}{6\epsilon_\sigma} \left(
{\cal E}^2-{\cal B}^2-2\gamma{\cal E}\cdot{\cal B}
\right) \,,
\end{align}
where the test field slow-roll parameter is defined as
\begin{align}
\epsilon_\sigma \equiv \dfrac{\dot{\sigma}^2}{2\Mp^2H^2} \,.
\end{align}
Demanding that the test field remains subdominant with respect to the inflaton field we require $\epsilon_\sigma < \epsilon_\phi$. Therefore we find that $R_{S}^\sigma > R_S$ so that the maximum value of $\xi$ allowed is even  less than in the case when the running field was the  inflaton field itself.  In other words, the backreaction is stronger for the test field $\sigma$ coupled to the gauge field. 

In summary, we need $\xi \lesssim 3$ to meet the backreaction condition. This is in contrast with the conclusion of 
\cite{Caprini:2014mja, Caprini:2017vnn} in which $\xi$ can take values 
in the range $\xi \sim 10-20$ by fine-tuning the value of $H/\Mp$ to a
very small value, say $H/\Mp \sim 10^{-20}$ to get the desired value of the observed magnetic fields. This value for Hubble parameter during inflation leads to $\epsilon_\phi \sim 10^{-32}$ which violates the backreaction constraint Eq.~(95) of \cite{Caprini:2017vnn} even for an inflaton field.

\subsection{Equilibrium state}


We see from Eq. \eqref{averageB} that the magnetic field experiences an equilibrium state in which the second term in the bracket falls off exponentially \cite{Talebian:2020drj}. Note that this is because  the magnetic field equation Eq. (\ref{langvinB2}) is in the form of OU stochastic differential equation where the classical drift term with a negative coefficient is balanced by the diffusion coefficient term. 

The time when the magnetic field reaches to its equilibrium is estimated as 
 \cite{Talebian:2020drj}
\begin{align}
\label{NeqB}
N^B_{\rm eq} \approx \dfrac{\ln 10}{n+2} \,,
\end{align}
when the exponential term \eqref{averageB} falls to less than $10^{-2}$. The amplitude of  each component of the dimensionless magnetic fields in the stationary state is given by
\begin{align}
\label{Beq}
\left<{\cal B}_i^2\right>_{\rm eq} &=
\frac{\DB^2}{2(n+2)} \, .
\end{align}

On the other hand, the situation for the electric field is very different as can be seen from Eq. \eqref{averageE}.  In the following we study the evolution of electric field in three different regimes, $\frac{1}{2}<n<2$, $n=2$ and $n>2$. 

\begin{enumerate}
	\item $\boldsymbol{\frac{1}{2}<n<2}$\\
	In this regime, the electric field does not grow with time and admits an equilibrium state. The time scale when the components of the electric field reach the equilibrium state is estimated as 
	\begin{align}
	N^E_{\rm eq} \approx \dfrac{2\ln 10}{2-n} \,,
	\end{align}
 with the equilibrium magnitude
	\begin{align}
	\left<{\cal E}_i^2\right>_{\rm eq} &=
	\gamma^2\left<{\cal B}_i^2\right>_{\rm eq}+\frac{(\DE +\gamma\DB)^2}{2(2-n)}
	-\frac{\gamma\DB(\DE +\gamma\DB)}{2}
	\nonumber
	\\
	&\simeq  \frac{\DE^2}{2(2-n)}\,.
	\label{Eeq}
	\end{align}
	Here we have neglected terms related to $\DB$ in favor of $\DE$ because  $\DB$ is smaller than  $\DE$ by a factor of $\varepsilon $ as seen from Eq.  \eqref{DXX}.
	
Since the electromagnetic fields falls into the stationary state for the parameter space $1/2<n<2$, one can use the alternative approach of probability distribution function to study the system.  This independent approach is studied  in App.~\ref{sec:PDF}.
	
	\item $\boldsymbol{n=2}$\\
	In this special case, the electric field becomes scale invariant, $i.e.$ $\DE$ does not depend on $\varepsilon$, and the evolution of the components of the  electric field is given by 
	\begin{align}
	\label{E_n_2}
	\dd {\cal E}_i &= 
	-\dfrac{\xi \DB}{\sqrt{2}}
	\, \dd N+D\textsub{E} \, \dd {\rm W}_i(N) \,.
	\end{align}
At the end of inflation, $N \simeq 60$, the condition \eqref{Constrant2} is violated for $\xi \gtrsim 3$. This is the hallmark of backreaction problems induced by the electric fields, as studied in previous literature using different approach.

	\item $\boldsymbol{n>2}$\\
In this case the electric field grows exponentially in time,
	\begin{align}
	\label{E_n>2}
	\left<{\cal E}_i^2(N)\right> &\simeq
	\dfrac{(\DE +\gamma\DB)^2}{2(n-2)}e^{2(n-2)N} \, \simeq \dfrac{H^2 \xi \, \sinh(2\pi \xi)}{6\pi^3\Mp^2(n-2)}
	\abs{\frac{ \Gamma (2n)}{2^n}}^2 \left(\dfrac{e^N}{\varepsilon}\right)^{2(n-2)}\, .
	\end{align}
Since $\varepsilon \ll 1\,, \xi \gtrsim 1$, there may be a very narrow band of the parameter space that the backreaction conditions \eqref{Constrant1} and \eqref{Constrant2} are satisfied initially. But then the condition \eqref{Constrant2} is violated and inflation is spoiled quickly. This again indicates the difficulties with the backreaction problem  in this setup
 induced by the electric fields.

\end{enumerate}

Hereafter, we only consider the range $\frac{1}{2}< n <2$ where both the  magnetic and electric fields experience equilibrium states with the amplitudes \eqref{Beq} and \eqref{Eeq} respectively. Also, as mentioned before, the amplitude of magnetic field scales with $\varepsilon$ like $\varepsilon^{n_B}$ while that of the electric field is stronger, scaling like  $\varepsilon^{n_B-1}$.
\\

Now let us have a closer look at the parameter $\varepsilon$. As we discussed in our  previous analysis \cite{Talebian:2020drj}, a lower bound on $\varepsilon$ can be found in the coarse-graining process
by considering  the longest wavelength observable on CMB scale, $k_{\rm CMB} \simeq 10^{-4}{\rm Mpc}^{-1}$. On the other hand, for the magnetogenesis mechanism, 
we are interested in the mode $k_{\rm Mpc}$ associated with the physical length scale $\sim {\rm Mpc}$ today. Then the smallest value for $\varepsilon$ is given by
\begin{align}
\varepsilon_{\rm Mpc} \simeq \dfrac{k_{\rm CMB}}{k_{\rm Mpc}} = 10^{-5} \,.
\end{align}
However, a larger value for $\varepsilon$ can be obtained by replacing $k_{\rm CMB}$ with the Planck observation's  pivot scale $k_* = 0.05~{\rm Mpc}^{-1}$, which results in $\varepsilon_{\rm Mpc} \simeq 10^{-2}$. Therefore we consider $\varepsilon_{\rm Mpc}$ in the range $10^{-2}-10^{-5}$ 
in the rest of the paper which is also small enough to meet the criteria of the long and short decomposition of stochastic analysis.

\subsection{Magnetic field at the end of inflation}

The stationary value of the dimensionless magnetic field  ${\cal B}$ is given in \eqref{Beq} which will be used to calculate the amplitude of the magnetic fields at the end of inflation. For simplicity, we assume an instantaneous reheating so we use the subscript ``${\rm rh}$" to indicate the corresponding value at the end of inflation. To proceed further, the characteristic properties of the primordial magnetic field must be translated into the stochastic language. By characteristic properties we mean the correlation scale $L_{\rm rh}$, the magnetic strength $B_{\rm rh}$ and the spectral index $n_B$ which, in the context of conventional approach,  are defined in \eqref{L}, \eqref{B_intensity} and \eqref{n_B}, respectively. The dictionary is as follows:

\begin{itemize}
	\item In stochastic approach, we deal with the coarse-grained magnetic field instead of the Fourier components. Therefore the magnetic strength \eqref{B_intensity} at the end of inflation is translated to
	\begin{align}
	B_{\rm rh} \equiv \sqrt{\langle B_{\rm l}^2 \rangle} \,,
	\end{align}
	in which $B_{\rm l} \equiv H \Mp \, {\cal B}_{\rm rh}$ is IR the part (long mode) of the magnetic field which is defined via the relation \eqref{Xsl}. Since for the parameter space $1/2<n<2$ the magnetic field components fall into the equilibrium state \eqref{Beq} the coarse-grained magnetic field at the end of inflation is given by
	\begin{align}
	\label{BendSto}
	B_{\rm rh} =  H \Mp \, \Big( \sum_{i=1}^{3}\left<{\cal B}_i^2\right>_{\rm eq}\Big)^{1/2}  \simeq \dfrac{\sqrt{3}}{\sqrt{2n+4}} H \Mp \DB \, ,
	\end{align}
	in which  the relation \eqref{Beq} has been used 
	for the component of the magnetic field in the equilibrium state. Now, using the value of $\DB$ given in Eq. (\ref{DB}), we obtain 
	$B_{\rm rh} \propto H^2  e^{\pi \xi}\varepsilon^{n_B}$.
	
\begin{figure}[t!]
	\centering
	\includegraphics[scale=0.8]{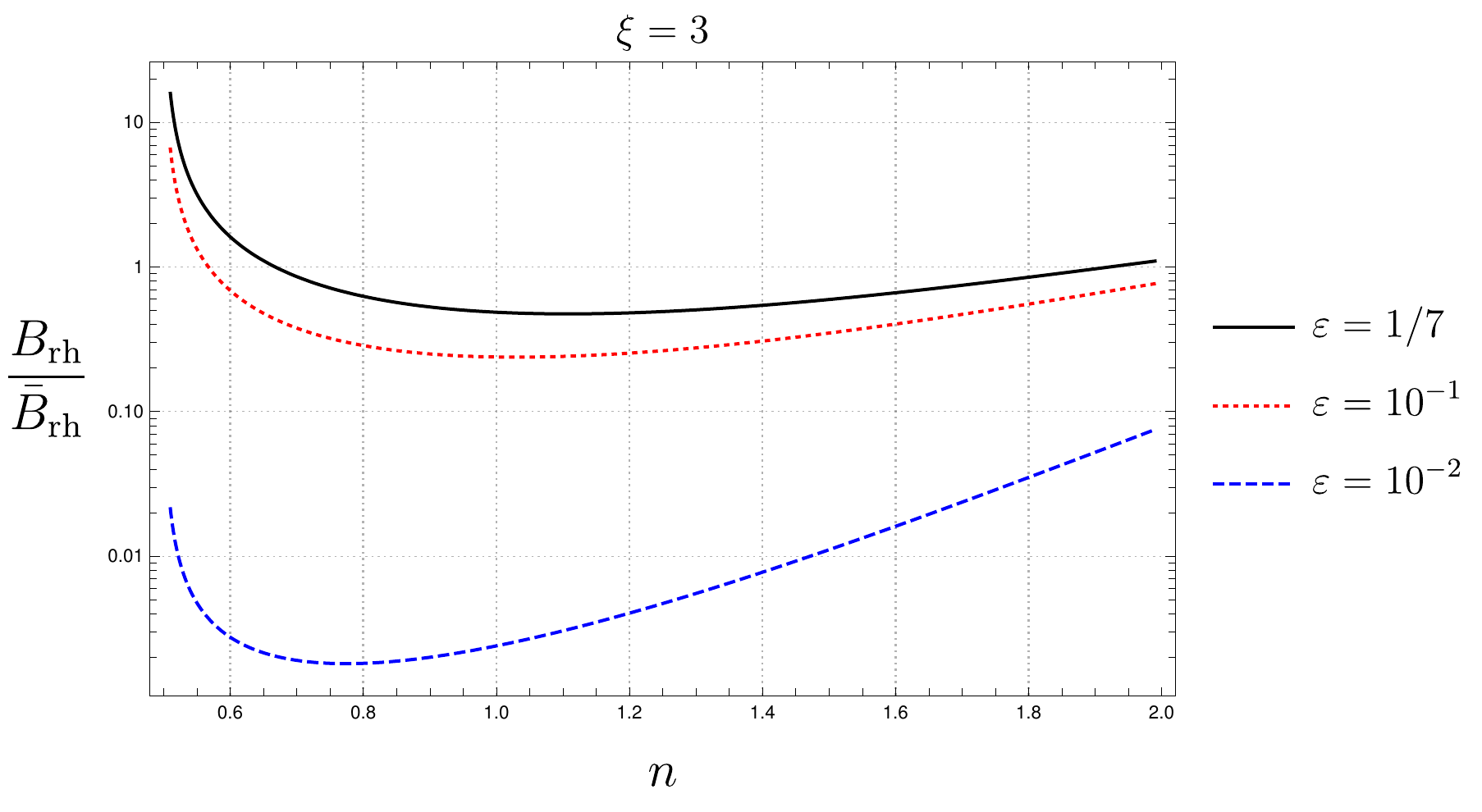}	
	\caption{\footnotesize The plot compares the amplitude of magnetic field 
	in  \eqref{B_rh} with  its counterpart  $\bar{B}_{\rm rh}$ in \eqref{brh} obtained  in the absence of stochastic effects.  We have chosen $\xi=3$ in order to avoid the back-reaction problem. For $\varepsilon \gtrsim {\cal O}(0.1)$ we have   $B_{\rm rh} \sim {\cal O}(\bar{B}_{\rm rh})$ but for the consistency of the stochastic formalism we require $\varepsilon \ll 1$. As a result  we find a smaller intensity for the magnetic field in the presence of stochastic noises.}
	\label{fig:Compare}
\end{figure}

	
	\item The correlation scale \eqref{L} can be used in stochastic approach as well. Therefore, integrating over the modes $\abs{k\tau} \lesssim \xi$, we obtain 
	\begin{align}
	L_{\rm rh} 	\simeq \frac{18\,\pi}{(3+2\,n)\,(5-2\,n)}\,\frac{\xi}{H}\,,
	\end{align}
	This scale is roughly given by the scale at which the power spectrum peaks.
	At the end of inflation, the correlation scale $L_{\rm rh}$ is much smaller than the smoothing scale $L_\varepsilon \sim 2\pi/(\varepsilon a H)$ by a factor of $\varepsilon/\xi$.
	\item The spectral index \eqref{n_B} is given by the power of parameter $\varepsilon$ in \eqref{BendSto}. Combining Eqs. \eqref{DXX} and \eqref{Beq}, we find $n_B = 3-n$ as mentioned in Eq. (\ref{nBb}). 
\end{itemize}
Therefore, the magnetic field intensity at the end of inflation is given by\footnote{The unit conversion, $1 \ {\rm GeV} = 3.8 \times 10^9 \ {\rm G}^{1/2}$, is used.}
\begin{align}
\label{B_rh}
B_{\rm rh} \simeq 5.3 \times 10^{55} \, {\rm G} \ \left(
\dfrac{H}{\Mp}
\right)^2 \,  \dfrac{\sqrt{\xi \sinh(2\pi\xi)} \, \Gamma(2n-1)}{2^n \sqrt{n+2}} \ \varepsilon^{n_B} \,.
\end{align}
The above expression  is the intensity of magnetic field at the end of inflation when the stochastic noises are taken into account. As seen, $B_{\rm rh}$ depends not only on the Hubble parameter during inflation $H$ but also it is a function of $n$ and $\xi$ for a fixed value of $\varepsilon$. As we  discussed before, the latter parameter controls the  scale dependency of magnetic field.

The expression \eqref{B_rh} can be compared with its counterpart Eq. 
\eqref{brh} which is obtained in the absence of the stochastic effects. In Fig.~\ref{fig:Compare}, we have plotted the ratio between these two amplitudes for the same values of $H$ and $\xi$ in term of $n$. The plot shows that in stochastic approach the  intensity of magnetic field at the end of inflation is smaller than what is obtained in the conventional method. This is because  the stochastic process controlling the dynamics of the magnetic field is an OU process in which the stochastic force is balanced by the frictional drift force. Therefore the tachyonic production of gauge field is controlled by the stochastic noises and the  amplitude of magnetic field becomes smaller than in conventional approaches where  no stochastic effects are included. 

\begin{figure}[t]
	\centering
	\includegraphics[width=0.8\linewidth]{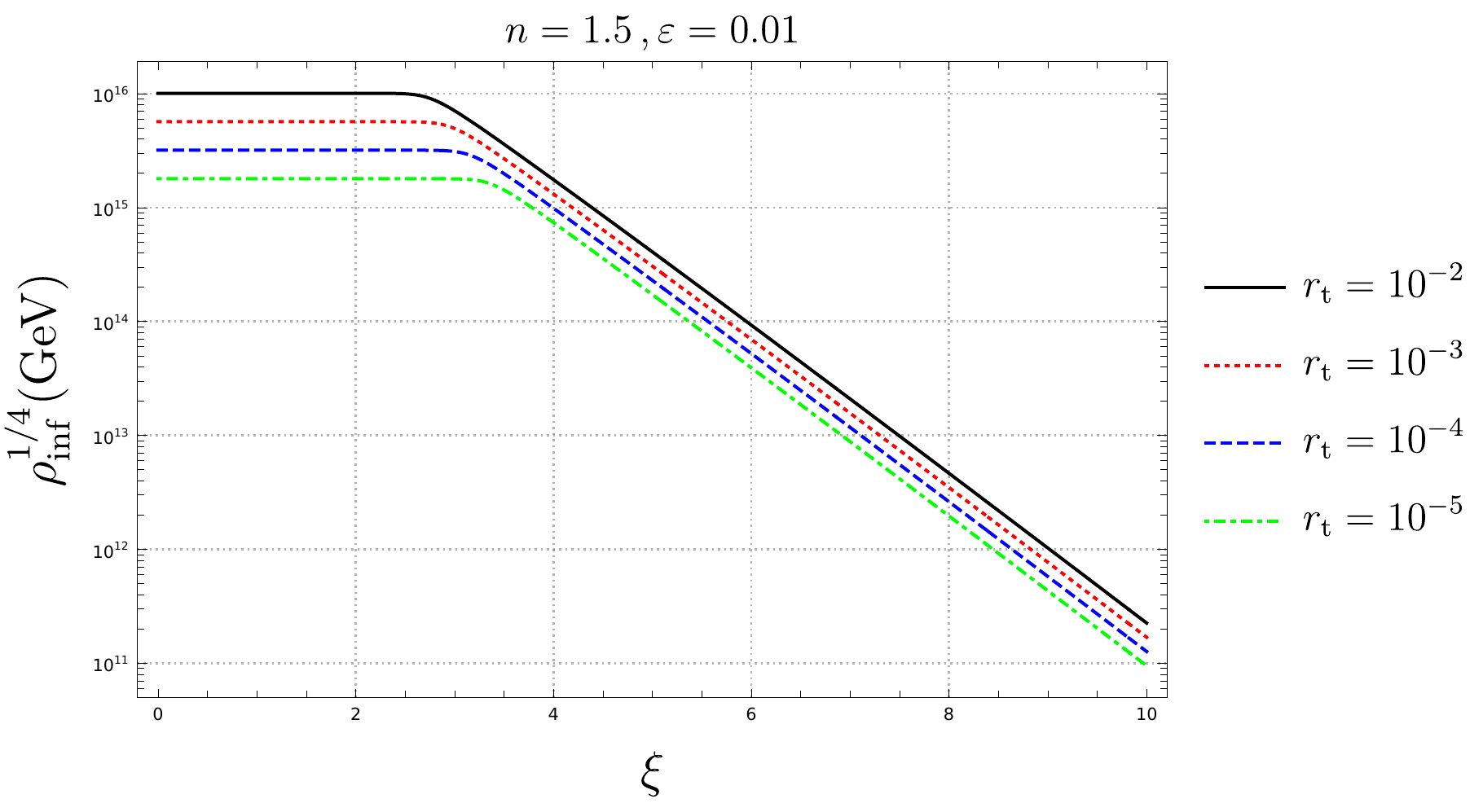}
	\caption{\footnotesize Inflationary energy scale as a function of the parameter $\xi$ with $n=1.5$ and  $\varepsilon=0.01$ for various tensor-to-scalar ratios. For $\xi  > 4$,  the  gauge field has the dominant contribution to $r_{\rm t}$ and inflation has a low energy scale,  $\rho_{\rm inf}^{1/4} \simeq 10^{-8}-10^{-15} \Mp$.}
	\label{fig:rho_inf}
\end{figure}

The amplitude of $H$ controls the energy scale of inflation and it is usually estimated by the tensor-to-scalar ratio $r_{\rm t}$. However, the tachyonic growth of the gauge field can source the tensor perturbations so there is a contribution from the electromagnetic source into $r_{\rm t}$ as well. In App.~\ref{GW}, we have estimated $r_{\rm t}$ in terms of the equilibrium amplitude of the electric field (for the interested range $1/2<n<2$) as a source for tensor perturbations. Based on the results from App.~\ref{GW} 
we have shown in Fig.~\ref{fig:rho_inf} how the energy scale of inflation depends on the model parameters.  Having the Hubble parameter during inflation \eqref{H_Mp} at hand, one finds the final result for the intensity of the magnetic field at the end of inflation $B_{\rm rh}(n, \xi , r_{\rm t})$. We comment that a very small value of $H/\Mp \sim 10^{-20}$ has been advocated in \cite{Caprini:2014mja,Caprini:2017vnn} in order to generate the observationally required value of the primordial magnetic field. As we mentioned previously, this is because they allowed for  $\xi \simeq 10-20$ while in our analysis, taking the backreaction and the stochastic effects all into account, we can go as far as  $\xi \simeq 3$.

In order to evaluate the intensity and the correlation length of the magnetic field today, we need to study its time evolution after the end of inflation. In App.~\ref{B_Evolution} we have reviewed the cosmological evolution of the helical and non-helical magnetic fields from the time of end of inflation, $B_{\rm rh}$,  until today $B_0$. In the next section we employ the results of App.~\ref{B_Evolution} to estimate the amplitude of the magnetic fields generated in the presence of the stochastic noises.

\section{Present Magnetic Fields}
\label{BToday}

A successful primordial magnetogenesis mechanism should generate an initial magnetic seed with a sufficient strength. Having \eqref{B_rh} as our primordial seed at the end of inflation, we study its evolution until today using the relations presented in App.~\ref{B_Evolution}. The seed field could be amplified by either astrophysical or primordial processes to produce the observed magnetic field today and to satisfy the IGMF and GMF observational constraints  as  pointed out in Introduction. In what follows, we will present these constrains with more details and then investigate the parameters of the model with which the constraints relating to GMF and/or IGMF could be satisfied.

\subsection{Observational constraints}
\label{obs}

We are interested in two classes of observations, GMF and IGMF, that may hint towards the primordial origin of the cosmological magnetic fields and could be sourced by the seed field \eqref{B_rh}. The former corresponds to galactic scale while the latter deals with the  extra-galactic scales. Although the astrophysical mechanisms of generation, such as the Biermann battery or the ejection of magnetic field from stars,  have  been assumed as possible seeds for the galactic dynamo and GMF, it would be difficult to provide fields that can account for the lower bound of IGMF because the bound applies in the absence of matter structure or ionized plasma \cite{Dolag:2010ni}. Therefore, IGMF observations can be considered as a strong hint on
the necessity of primordial magnetic seeds. The details of the observations are as follows.
\vspace{1cm}

\begin{enumerate}

\item \textbf{GMF:\\} 
	Using a number of techniques, magnetic fields of the order of
	\begin{align}
	B_{\rm GMF} \sim \mu{\rm G} \,,
	\end{align}
	are observed in Galaxies which are about tens of ${\rm Kpc}$. For example  our galaxy is permeated by a magnetic field with strength $3-4 \ \mu{\rm G}$ \cite{kronberg1994extragalactic} while magnetic fields with similar magnitudes (with strength $1-10 \ \mu{\rm G}$) have also been observed in cluster of galaxies on scales of up to $ \sim 0.1 {\rm Mpc}$ \cite{taylor1993magnetic,Eilek:1999yk}. A primordial magnetic seed with the minimal amplitude $\sim n{\rm G}$ can be amplified to desired strength by simple adiabatic contraction while the seeds with much smaller amplitudes must be amplified by stronger processes, \textit{e.g.} galactic dynamo mechanism. 
	
The dynamo mechanism transfers the kinetic energy of fluid into magnetic energy. More precisely, the coarse-grained hydrodynamics fluctuations can amplify a weak seed of magnetic field by providing the electromotive forces.\footnote{For comprehensive reviews of magnetic fields in the early Universe see for instance \cite{Giovannini:2003yn,Grasso:2000wj}.}. For instance, a field of order $10^{-30} \ {\rm G}$ at $10 \ {\rm kpc}$ is sufficient to initiate the dynamo process \cite{Davis:1999bt}. On the other hand, it is claimed in \cite{Giovannini:2003yn}  that a seed field of $10^{-23} \ {\rm G}$ at $\sim {\rm Mpc}$ is needed to initiate the dynamo mechanism. To estimate the intensity of magnetic field at $\sim {\rm Mpc}$  scale via inverse cascade process we follow Ref.~\cite{Caprini:2014mja,Brandenburg:2004jv} in which  a seed field in the range of
	\begin{align}
	\label{B-seed}
	10^{-23} {\rm G} \lesssim B_{\rm seed} \lesssim 10^{-21} {\rm G} \,,
	\end{align}
	at ${\rm Mpc}$ is required to explain the observed $\mu{\rm G}$ magnetic fields in Galaxies via dynamo mechanism. It must be noted that due to complicated galactic magnetohydrodynamics process, there is a large uncertainty on the ranges given in Eq.  \eqref{B-seed}. Moreover, the observation of magnetic fields with the same order in protogalactic clouds at high redshift is against the validity of galactic dynamo mechanism~\cite{Kronberg:2007dy,Bernet:2008qp}. However, it is usually assumed that the observed GMF to be the end product of this mechanism~\cite{Brandenburg:2004jv}. Therefore, we consider \eqref{B-seed} as a reference value for the seed amplitude.
	
\item \textbf{IGMF:\\}
IGMF constraint is based on the non-observation of GeV photons from TeV blazars and active galactic nuclei \cite{Neronov:1900zz}. IGMF leads to a lower bound on the intensity of magnetic fields in IGM with correlation length of Mpc or more~\cite{Vovk:2011aa}. This bound is strengthened for the smaller correlation length. 	Several observations  \cite{Ade:2015cva,Ade:2015cao,Pshirkov:2015tua,Giovannini:2017rbc,Sutton:2017jgr,Jedamzik:2018itu,Paoletti:2018uic,Bray:2018ipq} have constrained the strength of the cosmological magnetic fields in this class to be \cite{Fujita:2019pmi,Fujita:2014sna}
	\begin{align}
	B_{\rm IGMF} \gtrsim 10^{-16}{\rm G} \times
	\left\lbrace
	\begin{array}{lc}
	1   &L_{\rm B} \gtrsim 1 {\rm Mpc}\\
	\\
	\sqrt{\dfrac{1 {\rm Mpc}}{L_{\rm B}}} &L_{\rm B} \lesssim 1 {\rm Mpc}
	\end{array}\right.
	\label{B-bound}
	\end{align}
	In addition, there is an upper bound $B_{\rm IGMF} \lesssim 10^{-9}{\rm G}$ coming from the CMB data. There are two points here that must be mentioned. First, the above observation puts a constrain not only  on the intensity of the magnetic field but also on its correlation length. 
	Second, the lower bound $10^{-16}{\rm G}$ on ${\rm Mpc}$ scales is not very rigid and it could take a wide range with width of three order of 
	magnitudes, $10^{-15}-10^{-18} {\rm G}$, depending on the details of  cascade emission and its time delay \cite{Taylor:2011bn,Vovk:2011aa}. 
\end{enumerate}

The large correlation length involved in the bounds \eqref{B-bound} and \eqref{B-seed} may hint towards the primordial origin of the cosmological magnetic fields.  We are interested in the parameter space in which the primordial magnetic fields \eqref{B_rh}  satisfy the lower bound in IGMF \eqref{B-bound} and provides the seed \eqref{B-seed} for GMF  assuming the galactic dynamo mechanism as the amplification mechanism.

In order to study the cosmological evolution of the primordial magnetic fields, we should consider the flux conservation as well as helicity conservation. The relations between the present amplitude of the magnetic field $B_0$ and the reheating value $B_{\rm rh}$ for the flux and helicity conservations are given in Eqs.  \eqref{B0nonhelical} and \eqref{B0helical} respectively. Depending on the characteristics of the magnetic field produced during  inflation as well as the characteristics of the environment through which it passes, one of the two conservation laws could be at work.  The modes that exit the horizon during inflation at the early times will be less affected by the plasma turbulence when they re-enter horizon after recombination. Therefore, the flux conservation is a good approximation to study their evolution. On the other hand, the modes which exit the horizon at the later time will encounter the turbulent plasma at the re-entry time. Hence their flux is not conserved so we study their evolution via helicity conservation. More precisely, there exits a special scale $k_{\rm diss}$ at which the Reynolds number of plasma is at the order of unity. Modes with $k> k_{\rm diss}$ ($k< k_{\rm diss}$) come across the plasma at the turbulent (viscus) regime so in order to study their subsequent evolution we can easily consider helicity (flux) conservation. In the following, we use the superscripts $F$ and $H$ to denote the present magnetic fields $B_0^{\rm F}$ and $B_0^{\rm H}$, which are evolved via the flux and helicity conservation respectively.

For a small value of $\xi$, the generated magnetic field is non-helical so one can track the evolution of the magnetic field by imposing the flux conservation. For helical fields, however,  the situation is different.  It is well-studied  that during the radiation dominated epoch, the helical magnetic fields undergo the inverse cascade process \cite{Banerjee:2004df, Campanelli:2013iaa}. During this process the intensity of the magnetic field decreases in the comoving frame
and its correlation scale increases while the  power is transferred from small to large scales. Furthermore,  it displays a property of self-similarity, \textit{i.e.} the magnetic spectral index at scales larger than the correlation scale remains unchanged.

\subsection{Flux conservation}
\begin{figure}[t!]
	\centering
	\includegraphics[scale=0.8]{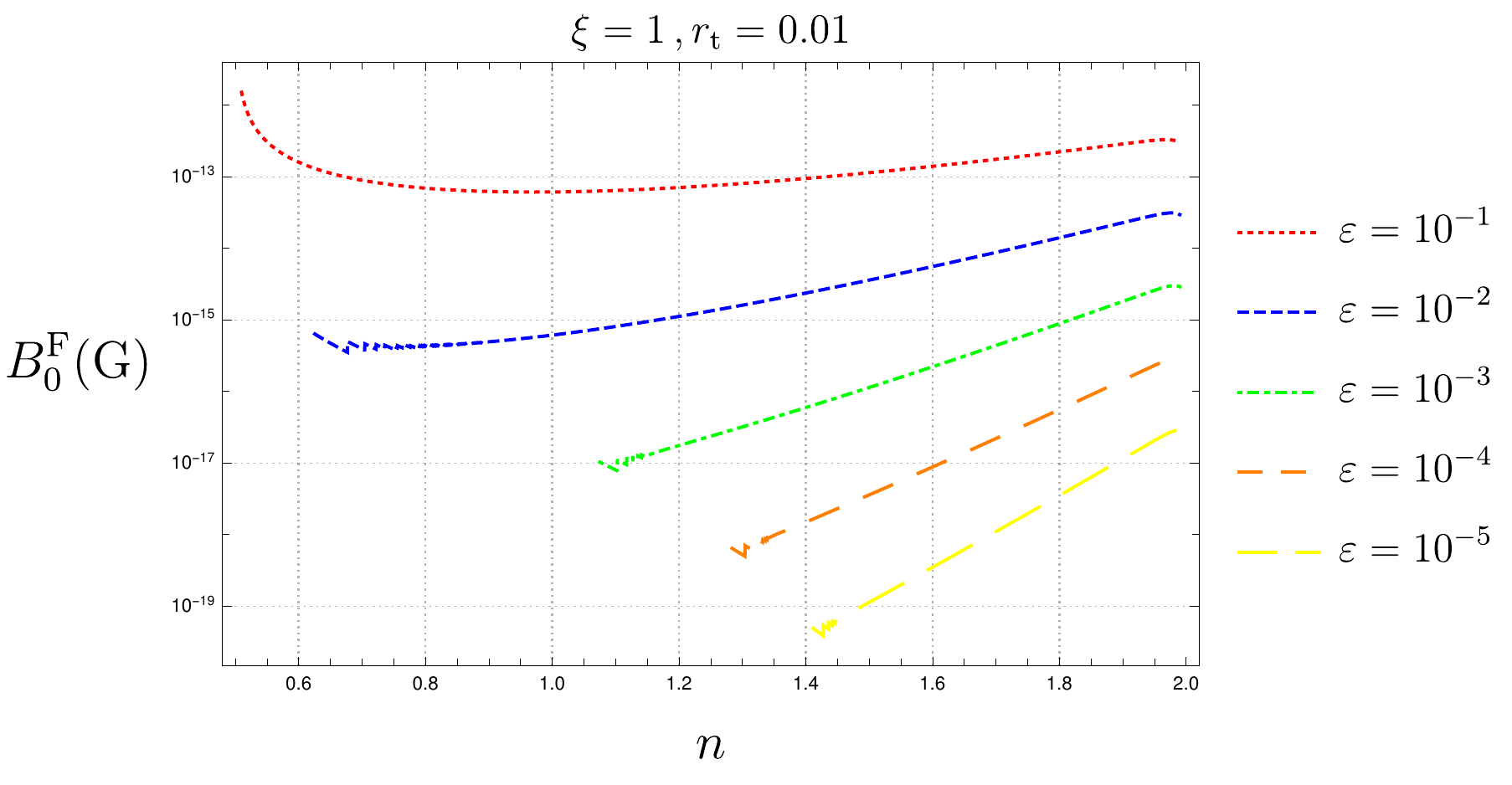}	
	\caption{\footnotesize Present value of magnetic field at Mpc scale in term of parameter $n$ according to \eqref{Bnow_nonhelical} for $\xi =1$
in which the flux conservation is assumed. As seen the generated magnetic field easily satisfies the observational IGMF bound \eqref{B-bound} while not the GMF bound via the galactic dynamo mechanism. One may consider a process weaker than the dynamo mechanism to amplify the intensity of order $10^{-15}{\rm G}$ to $\mu{\rm G}$ observed in Galaxies.}
	\label{fig:B0_radiation}
\end{figure}

For $\xi \lesssim 1$ the magnetic field is basically non-helical. Neglecting the resistivity and the turbulence of the primordial plasma, one can estimate the strength of the magnetic fields at the present time. Assuming the radiation-like dilution for the electromagnetic energy density and also an instant reheating scenario after the end of inflation\footnote{See \cite{Kobayashi:2019uqs} for a different discussion.} leads to
\begin{align}
\label{Bnow_nonhelical}
B_0^{\rm F} \simeq 
 1.7 \times 10^{-6} \, {\rm G} \ \left(
 \dfrac{H}{\Mp}
 \right) \,  \dfrac{\sqrt{\xi \sinh(2\pi\xi)} \, \Gamma(2n-1)}{2^n \sqrt{n+2}} \ \varepsilon^{n_B}
\,,
\end{align}
in which we have inserted \eqref{B_rh} into \eqref{B0nonhelical}. Using the relation \eqref{H_Mp} for the Hubble parameter, which represents the energy scale of inflation, the behaviour of the present amplitude of the magnetic field $B^{\rm F}_0$ in terms of the parameters of the model $n, ~ \xi,~ r_{\rm t}$ is obtained. The result is plotted in Fig.~\ref{fig:B0_radiation}. The energy scale of inflation  is also plotted in Fig.~\ref{fig:rho_inf}.
We see that taking into account the stochastic noises,  an acceptable amplitude for the present magnetic field is generated \cite{Talebian:2020drj}
with a high energy scale of inflation. 

\begin{figure}[t!]
	\centering
	\includegraphics[scale=0.8]{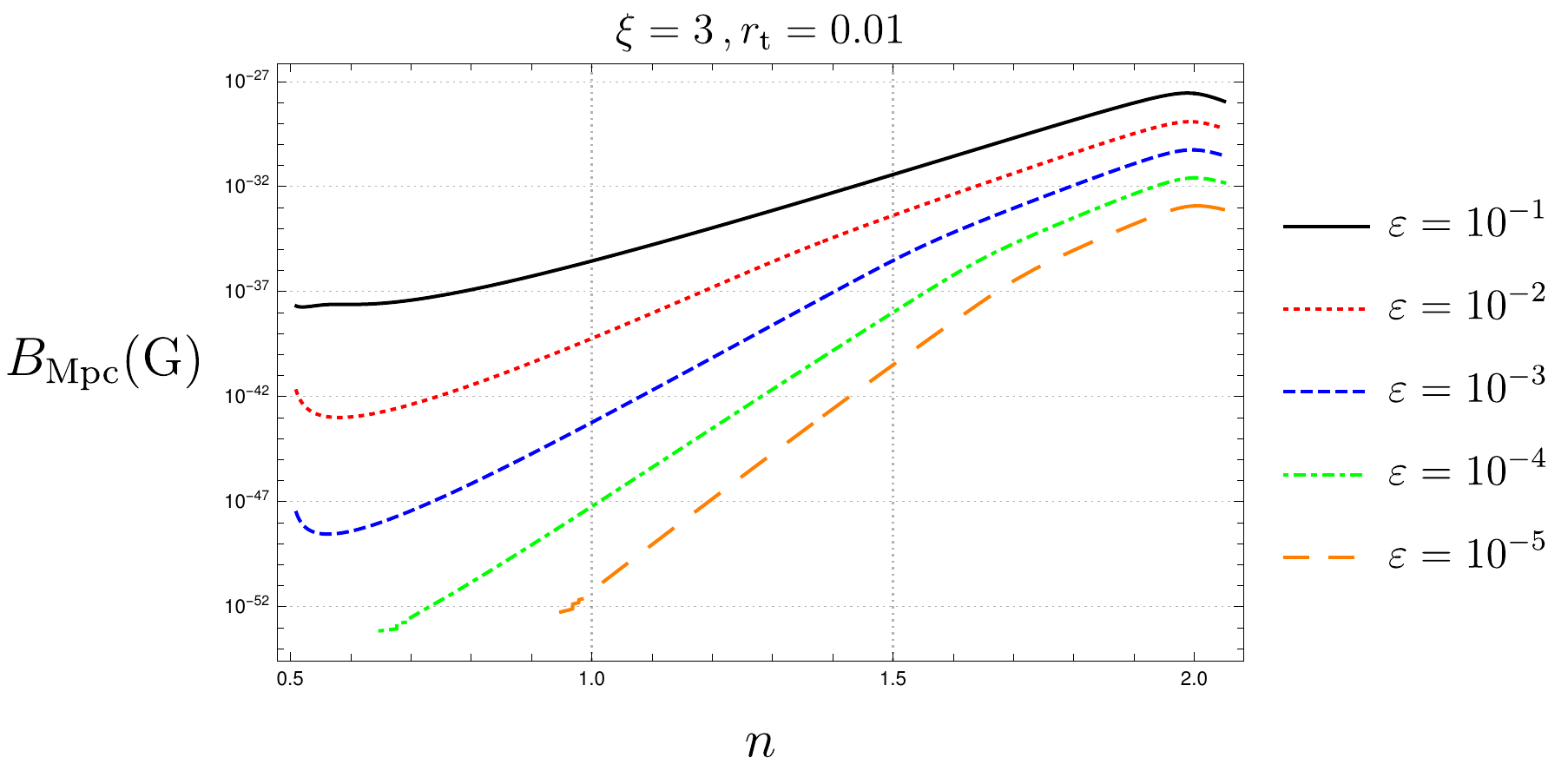}	
	\caption{\footnotesize Present value of magnetic field at Mpc scale in term of parameter $n$ according to \eqref{BMpc-helical} in which the helicity conservation is assumed for $\xi =3$. As seen the primordial seed field \eqref{B_rh} is not strong enough to provide the initial seed needed in \eqref{B-seed} for the dynamo mechanism. One will need  a process stronger than the dynamo mechanism to amplify the intensity of order $10^{-30}{\rm G}$ to $\mu{\rm G}$ observed in Galaxies for the parameter space $n \lesssim 2$.}
	\label{fig:B0_H_Mpc}
\end{figure}

\subsection{Helicity conservation}

For $\xi >1$, the generated magnetic
field at the end of inflation is maximally helical\footnote{ For a different mechanism of helical magnetogenesis see \cite{Kushwaha:2020nfa}. }. 
After inflation, the thermal cosmic plasma contains many relativistic charged
particles and can be treated as an MHD plasma. In this
limit the electric field is damped away and the magnetic field undergoes an inverse cascade due to
helicity conservation. Therefore the discussion presented in App.~\ref{helicity} is relevant.

Inserting Eq. \eqref{B_rh} into Eq. \eqref{B0helical},  the intensity of the present magnetic field at the correlation scale $L_0^H =10^8 {\rm Mpc} \left(B_0^H/{\rm G}\right)$ is obtained to be 
 \begin{align}
\label{B0H}
B_0^{\rm H} &= 5 \times 10^{-16} {\rm G} \ \left( \dfrac{H}{\Mp}
\right)^{1/2} \
\left(
\dfrac{\Gamma^2(2n-1)}{2^{2n+1}(n+2)}e^{2\pi\xi}\varepsilon^{2n_B}
\right)^{1/3} \,.
\end{align}
 Since we are interested in ${\rm Mpc}$ scales, the correction arising from the scale dependence must be taken into account via the relation \eqref{B0ell}. Doing so, the amplitude of magnetic field at ${\rm Mpc}$ scales is obtained to be  
\begin{align}
\label{BMpc-helical}
B_{\rm Mpc} = B_0^{\rm H}\left(\dfrac{L_0^{\rm H}}{{\rm Mpc}}\right)^{n_B} \,.
\end{align}

\begin{figure}[t!]
	\centering
	\includegraphics[scale=0.6]{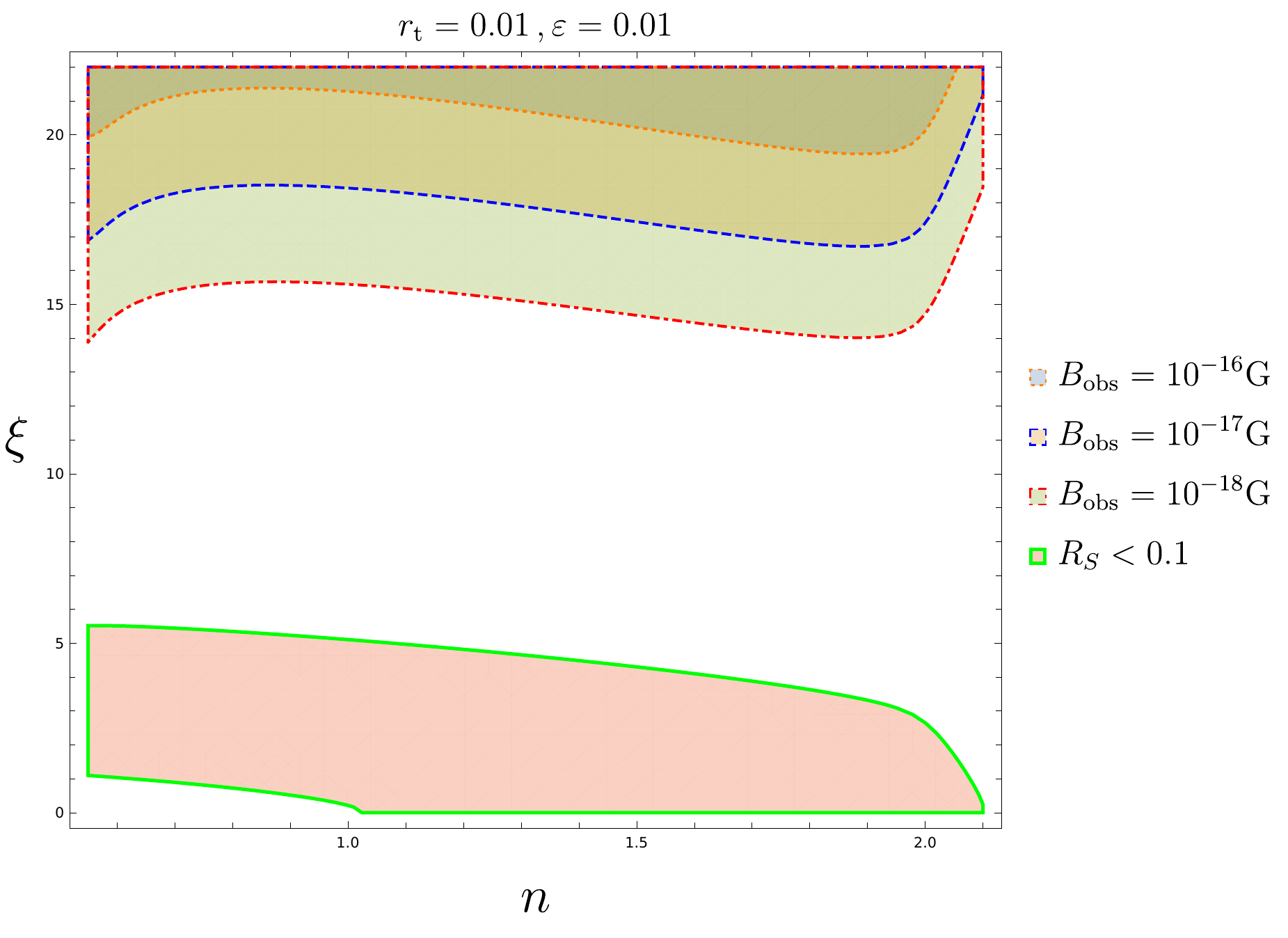}	
	\caption{\footnotesize The allowed region with small backreaction 
	 $R_S<1$ \eqref{Constrant2} is confined  in the lower bounded area (red area).  The parameter space where  the IGMF constraint \eqref{constraint} is satisfied is confined in the upper parts bounded by different curves (red, blue and orange). There is no overlapping area, which means that the model  with the maximally helical seeds 
	 can not satisfy the  IGMF constraint.}
	\label{fig:Bobs-RS}
\end{figure}

In Fig.~\ref{fig:B0_H_Mpc}, we have presented $B_{\rm Mpc}$ in terms of $n$ and found that the maximum value of the intensity of magnetic field is not stronger that $10^{-28}{\rm G}$ which is too small to be considered as a seed field \eqref{B-seed} to initiate the galactic dynamo and explain the GMF constraints. One would need a stronger process in order to amplify this small seed value to the desired amplitude\footnote{{It should be noted that we have considered the strong coupling regime in which $0<n<\frac{1}{2}$ as well and the result for the magnetic field didn't change much from Fig. \ref{fig:B0_H_Mpc}. }}.

To investigate the IGMF constraint, we follow the method used in \cite{Caprini:2015gga,Caprini:2017vnn}.  The magnetic fields  with the minimal amplitude of the order of $10^{-18} ~ {\rm G}$ and correlation length $L_0^{\rm H} \gtrsim D_{\rm e}$ can explain the non-observation of GeV gamma-ray cascades around blazars in the Inter Galactic Medium (IGM)
~\cite{Neronov:1900zz,Taylor:2011bn,Vovk:2011aa} in which  $D_{\rm e}$ is the electron/positron energy loss length for inverse Compton scattering. The correlation length of our setup is typically in the range $L_0^H \simeq 10^{-3}~{\rm pc} - 10^{-9}~ {\rm pc}$ which is much smaller than $D_{\rm e} \simeq 80 ~{\rm Kpc}$. Taking into account the correction
arising from the scale dependence for $n_B>1/2$ ( equivalent to $n<5/2$), the constraint \eqref{B-bound} for $L_0^H \ll D_{\rm e}$ 
is translated into the following upper bound~\cite{Caprini:2015gga}
\begin{align}
\label{constraint}
B_0^{\rm H} ~ \gtrsim ~B_{\rm obs}~ \sqrt{\dfrac{D_{\rm e}}{L_0^{\rm H}}} ~ \sqrt{\dfrac{10n_B-5}{n_B}} \,,
\end{align}
where $B_{\rm obs} = 10^{-18} {\rm G} -10^{-16} {\rm G} $. 

The above constraint will need a large value of $\xi$, say $\xi \sim 10-20$. However, this large value of $\xi$ is not allowed since it induces large backreactions on the scalar field dynamics.  In Fig.~\ref{fig:Bobs-RS} we have shown the allowed regions in parameter space $\xi-n$ in which the back-reaction problem is bypassed and the constraint \eqref{constraint} is satisfied. As seen, there is no overlapping  region and the primordial seed fields \eqref{B_rh}  can not satisfy the IGMF constraint \eqref{B-bound}.


\section{Conclusion}
\label{Conclusion}

In this paper, we have revisited the mechanism of magnetogenesis in the $I^2 F \tilde{F}$  inflationary model by taking into account the stochastic effects of the electromagnetic fields perturbations.  We have derived the associated Langevin equations for the electric and magnetic fields and have calculated their two points correlations. The corresponding Langevin equations are in the form of  Ornstein-Uhlenbeck stochastic differential equations with a negative drift coefficient. 
We have shown that both the electric and magnetic fields settle in equilibrium states  very soon  with the strengths given by Eqs. \eqref{Eeq} and \eqref{Beq}, respectively.

We also checked the back-reaction constraint and found that the instability parameter has an upper bound $\xi \lesssim 3$ in order for the backreaction of electromagnetic fields on the dynamics of inflaton field to be under control.  This bound is consistent with the results of \cite{Durrer:2010mq, Salehian:2020asa}. In addition, the backreaction effects become stronger when we use a test field instead of inflaton because the slow-roll parameter associated with the test field is smaller than that of  inflaton.

The results shows that the stochastic effects cause   the amplitude of magnetic field at the end of inflation to be smaller than what is obtained in conventional method by at least two order of magnitudes; see Fig.~\ref{fig:Compare}.
The stochastic forces tame the tachyonic growth of IR modes which are described by an OU-type stochastic differential equation. The process settles the fields into equilibrium states and decrease their exponential growths.

The setup with  $\xi >1$ produces magnetic fields with a net helicity. Therefore, the helicity conservation must be considered for the evolution of magnetic field from the end of inflation until today. But, as mentioned above, the backreaction constraint requires  $\xi \lesssim 3 $ so the tachyonic growth of the electromagnetic perturbations are limited and  
the observational constraints \eqref{B-seed} and \eqref{B-bound} are not satisfied. Therefore in the parameter space where the backreaction is under control  the model  is not able to provide a chiral primordial seed for GMF and IGMF. Furthermore,  as shown in Fig.~\ref{fig:rho_inf}, 
with  $\xi \lesssim 3$,  the energy scale of inflation can be as high as  $10^{-3} - 10^{-4} {\Mp}$. These results are in contrast with the results of \cite{Caprini:2014mja, Caprini:2017vnn} in which it is claimed that the model with $\xi \sim {\cal O}(10)$ is able to account not only for the IGMF observations but also to initiate the galactic dynamo by providing the seed field in the range \eqref{B-seed} while inflation is happening at low energy scale. 

On the other hand, for $\xi<1$, the generated magnetic field is not helical and one can simply study the evolution of the magnetic field via flux conservation. This yields the present magnetic field with the amplitude $B_0^F \simeq 10^{-13}{\rm G}$ on ${\rm Mpc}$ for $n \simeq 2$ which is 
well suitable into the IGMF bound \eqref{B-bound}. The generated seed field is too strong for the galactic dynamo mechanism but one can consider another weaker processes, \textit{e.g.} adiabatic contraction, to amplify these magnetic fields to provide the intensity of order of $\sim \mu {\rm G}$ on galactic scales.

\vspace{0.7cm}

{\bf Acknowledgments:}  H. F. and A. T. would like to thank the ``Saramadan" federation of Iran for the partial supports.  

\appendix

\appendix

\section{Cosmological evolution of magnetic fields}
\label{B_Evolution}
In this appendix, we briefly review the evolution of cosmological magnetic fields in an expanding Universe filled with and without the cosmic plasma.  We refer the reader to \cite{Durrer:2013pga,Banerjee:2004df} for the detailed and critical reviews of the literature on the subject.

Due to the homogeneity and isotropy of the Universe, it is more convenient to study the properties of magnetic
field in terms of its Fourier components,
\begin{align}
{\bf B}({\bf k},t) = \int \dd^3x \ {\bf B}(\bfx,t) \ e^{i {\bf k}\cdot \bfx} \,.
\end{align}
Using Eqs. \eqref{electric} and \eqref{A_tilde}, we obtain
\begin{align}
\label{B_k}
B_i({\bf k},\tau)=\sum_{\lambda=\pm} 
 e_i^\lambda({\bf k}) 
\bigg(
B_\lambda(k,\tau) \, a_{{\bf k},\lambda} + B^*_\lambda(k,\tau) \,  a^\dagger_{-{\bf k},\lambda}
\bigg) 
\,,
\end{align}
where $k = \abs{\bf k}$ and $B_\lambda(k,\tau) = \lambda \ k \ v_{k,\lambda}(\tau)$. The spatial structure of magnetic fields is statistically the same at any location in the Universe which implies  that the expectation values of the magnetic fields only depend on $\bf k$, $\delta_{ij}$ and $\epsilon_{ijk}$ as well as their combinations of these. 

 The two-point function of the Fourier components of the magnetic field, which is a divergence-free vector field, in the comoving coordinate can be written as
\begin{align}
\label{B_power}
\langle B_i({\bf k},\tau)\,B_j^*({\bf q},\tau)\rangle = 
(2\pi)^3\,\frac{\delta({\bf k}-{\bf q})}{k^3} \  \left(
(\delta_{ij}-\hat{k}_i\hat{k}_j)\,{\cal P}_B(k,\tau) - i \, \epsilon_{ijl} \, \hat{k}_l \ {\cal H}_B(k,\tau)
\right)\, ,
\end{align}
where $\hat{\bf k}={\bf k}/k$ and the  bracket $ \langle \  \rangle$ denotes an ensemble average. 

The symmetric and antisymmetric part of the above correlation 
are denoted by ${\cal P}_B$ and ${\cal H}_B$ respectively, \textit{i.e.}
\begin{align}
\sum_{\lambda=\pm}^{} \langle B_\lambda({\bf k},\tau)\,B_\lambda^*({\bf q},\tau)\rangle &=(2\pi)^3\,\frac{\delta({\bf k}-{\bf q})}{k^3} {\cal P}_B(k,\tau) \,,
\\
\sum_{\lambda=\pm}^{} \lambda \, \langle B_\lambda({\bf k},\tau)\,B_\lambda^*({\bf q},\tau)\rangle &=(2\pi)^3\,\frac{\delta({\bf k}-{\bf q})}{k^3} {\cal H}_B(k,\tau) \,.
\end{align}
The symmetric part of the spectrum determines the energy density, \begin{align}
\rho_B(\tau) \equiv 
\dfrac{1}{2\pi^2} \int \dd \ln k \ {\cal P}_B(k,\tau) \,.
\end{align}
Therefore ${\cal P}_B(k,\tau)$ is related to the magnetic energy density per logarithmic wave number via
\begin{align}
\label{P_B}
{\cal P}_B(k,\tau) = 2\pi^2 \ \dfrac{\dd \rho_B(k,\tau)}{\dd \ln k} \,.
\end{align}
The magnetic helicity is defined as 
\begin{align}
H(V,\tau) = 
\int_V \dd^3x \ \langle {\bf A}(\bfx,\tau) \cdot {\bf B}(\bfx,\tau) \rangle \,,
\end{align}
where $V$ is a volume through the boundary of which no magnetic field lines cross. 
For the gauge in which the 3D vector potential $\bf A$ is transverse, we define the magnetic helicity density as
\begin{align}
\mathpzc{h}(\tau) \equiv \langle {\bf A}(\bfx,\tau) \cdot {\bf B}(\bfx,\tau) \rangle =\dfrac{1}{2\pi^2} \int \dd \ln k \ {\cal H}_B(k,\tau) \,,
\end{align}
Hence ${\cal H}_B(k,\tau)$ is related to the helicity density per logarithmic  interval,
\begin{align}
{\cal H}_B(k,\tau) = 2\pi^2 \ \dfrac{\dd \mathpzc{h}(\tau)}{\dd \ln k} \,.
\end{align}

It is convenient to assign two characteristic properties to the magnetic fields. First,  the characteristic correlation length 
$L$, which is sometimes called the correlation scale, is defined via \begin{align}
\label{L}
L \equiv \dfrac{\int \dd \ln k \ \left(
	\frac{2\pi}{k}
	\right){\cal P}_B(k)}{\int \dd \ln k \ {\cal P}_B(k)} \,,
\end{align}
which is a measure of the scale  containing most of the magnetic energy. 

Second, the scale-averaged magnetic strength is given by
\begin{align}
\label{B_intensity}
B \equiv \sqrt{2 \rho_B} \,,
\end{align}
while the characteristic magnetic field strength
at scale $\ell =2\pi/k$ is defined as
\begin{align}
B_\ell \equiv \sqrt{2 \dfrac{\dd \rho_B}{\dd \ln k}} \Bigg|_{k=2\pi/\ell} = \dfrac{\sqrt{{\cal P}_B(k)}}{\pi}\Bigg|_{k=2\pi/\ell} \,.
\end{align}
In addition, the magnetic spectral index on large
scales is defined as
\begin{align}
\label{n_B}
n_B \equiv \dfrac{1}{2} \dfrac{\dd \ln {\cal P}_B(k)}{\dd \ln k} \,.
\end{align}
For example in the model $I(\tau)F_{\mu \nu}F^{\mu \nu}$ with $I(\tau) \propto \tau^n$ the magnetic field spectral index is given by $n_B=\frac{5}{2} - | n- \frac{1}{2}|$ so that the cases with $n=3$ and $n=-2$ lead to scale invariant magnetic spectra. 

To study the expected relic magnetic field, which might survive until
the present epoch, we must have enough knowledge about the initial spectrum of the magnetic field generated during inflation, i.e. ${\cal P}_B$ and ${\cal H}_B$, and know their evolution well after inflation and reheating phase. 

In the first approximation, the conductivity of the Universe, which is very high after reheating, must be considered. Therefore any electric fields produced during inflation will be damped very rapidly after inflation while the magnetic field is frozen. This is why we consider magnetogenesis models prior to reheating. The electric conductivity converts the generated electromagnetic modes into a frozen magnetic field which obeys adiabatic dilution, \textit{i.e.} the magnetic power spectrum decays as ${\cal P}_B \propto a^{-4}$ after the mode function is frozen due to electric conductivity.


A better approximation is to consider a plasma environment instead of an electrically conductive medium. Due to the presence of many relativistic charged particles after inflation, the thermal cosmic plasma can be treated as a Magneto Hydro Dynamic (MHD) plasma. An important characteristic of the fluid flow is given by its local kinetic Reynolds number, denoted by $R_{e}$. The Reynolds number is a measure of the relative importance of fluid dissipative terms in the Euler equations of MHD fluid. In general, in the MHD limit the electric fields are damped away while the magnetic fields evolution must be studied in two different regimes; the turbulent regime, when $R_{e} \gg 1$, and the viscous regime, when $R_{e} \ll 1$. In the former regime, the magnetic field is damped on small scales which leads to a maximally helical field, \textit{i.e.} one of the polarization modes vanishes \cite{Banerjee:2004df
}. Therefore, the magnetic fields undergo an inverse cascade due to \textit{helicity conservation} \cite{Brandenburg:2004jv}. This effect is active as long as $R_e >1$ on the scale under consideration. Therefore, the fluid is turbulent in the regime in which 
the decay rate of the total energy only depends on the flow properties on the integral scale and is independent of dissipative terms. This regime is applicable well before the neutrino decoupling and recombination. After the end of the turbulent phase, magnetic fields are damped on small scales by viscosity and evolve by \textit{flux conservation}, so that $B \propto a^{-2}$ on large scales.

In the viscous regime, the decay of magnetic energy depends on the magnitude of viscosities. This regime describes the state of the cosmic plasma, before recombination, on the scales smaller than the damping scale $k_{\rm diss}$ at which $R_e$ becomes of order unity. Both the turbulent motion of the fluid and the magnetic field are damped exponentially by viscosity. Furthermore, there is the effect of ambipolar diffusion after recombination, when the Universe is a weakly ionized fluid. This diffusion is due to the ion-neutral mixture in the tightly coupled regime which inserts an additional dissipative term in the MHD equations. We refer the reader to the references \cite{Banerjee:2004df,Durrer:2013pga} for more details about the general features of the evolution of magnetized fluids, such as the decay of energy density as well as the growth of magnetic field coherence length, in the turbulent and viscous regimes. 
 

In what follows, we will study the evolution of the magnetic fields from the end of inflation (reheating) till the present time in terms of two different assumptions: flux conservation ($\rho_B \propto B^2 = const.$) and helicity conservation ($\mathpzc{h} \propto B^2 L = const.$). For the expanding Universe these two conservation laws leads to $B^2 \propto a^{-4}$ and $B^2 L \propto a^{-3}$, respectively.

\subsection{Flux conservation}
\label{flux}

In order to estimate the strength of the magnetic fields at the present time 
we assume the radiation-like dilution for the electromagnetic energy density and neglect the high conductivity and turbulence of the primordial plasma with   an instant reheating scenario (see \cite{Kobayashi:2019uqs} for a controversial discussion). Due to the flux conservation, the amplitude of magnetic field at the present time, denoted by $B_0$, is given by
\begin{align}
B_0 =  \left(\dfrac{a_{\rm rh}}{a_0}\right)^2 B_{\rm rh} \,,
\label{B-now}
\end{align}
where the amplitude of magnetic field at the end of inflation is denoted by
$B_{\rm rh}$ and $a_{\rm rh}$ and $a_0$ are the values of the scale factor at the end of inflation and at present, respectively. To simplify the situation further, 
we assume that the Universe was radiation dominated throughout its history with a reasonable  accuracy. Then, we have
\begin{align}
\dfrac{a_0}{a_{\rm rh}} = \left(
\dfrac{g^*_{\rm rh}}{g^*_0}
\right)^{1/3} \dfrac{T_{\rm rh}}{T_0} \,,
\label{a-scale}
\end{align}
in which $g^*_0$ and $g^*_{\rm rh}$ are the effective numbers of relativistic degrees of freedom at the present time and at the time of reheating respectively.  Moreover, an instant reheating scenario allows to express $T_{\rm rh}$ in terms of Hubble rate at the end of inflation $H$ as
\begin{align}
\label{T_rh}
T_{\rm rh} \simeq 1.5 \times 10^{31} \ {\rm K}\left(
\dfrac{g^*_{\rm rh}}{106.75}
\right)^{-1/4}\left(\dfrac{H}{\Mp}\right)^{1/2} \, .
\end{align}
Using these relations, the amplitude of the observed magnetic field at the present time is given by
\begin{align}
\label{B0nonhelical}
B_0 \simeq 3.2 \times 10^{-62}  
\left(\dfrac{H}{\Mp}\right)^{-1} 
~B_{\rm rh} \, ,
\end{align}
where  
we have set $g^*_0=3.36$ and $g^*_{\rm rh}=106.75$.

\subsection{Helicity conservation}
\label{helicity}

The evolution of $B$ and $L$ during post-inflationary epoch undergo several different phases: turbulent, viscous and free-streaming \cite{Banerjee:2004df, Durrer:2013pga}. Not only the initial values of the intensity and the correlation scale determine what phase the Universe is, but also particle species with the longest mean free path (neutrinos, followed by photons after neutrino decoupling) have significant effects on the evolution of the magnetic energy via the temperature of the kinetic viscosity of the plasma.

It is well-known that in the turbulent fluid with non vanishing helicity the helical magnetic field undergoes a process known as inverse cascade during the radiation dominated epoch. Considering this process, the comoving $L$ increases and its comoving intensity $B$ decreases. The power is transferred from small scales to large scales while the magnetic spectrum at scales larger than $L$, \textit{i.e.} $\ell > L$ maintains its spectral index unchanged. 
Consequently, the amplitude of the magnetic field on large scale is given by
\begin{align}
\label{B0ell}
B_{\ell > L} = B \left(\dfrac{L}{\ell}\right)^{n_B} \,,
\end{align}
displaying a property of self-similarity \cite{
	Banerjee:2004df}. It must be noted that the inverse cascade is not effective in the case of a (nearly) scale invariant spectrum even for a fully helical magnetic fields~\cite{Brandenburg:2017rcb}.

Taking into account high conductivity along with the turbulence of the primordial plasma, the magnetic fields evolve conserving (comoving) magnetic helicity density (instead of magnetic flux) via the inverse cascade process. Therefore, in an expanding Universe,  we have
\begin{align}
\label{B_0_1}
B_0^2 L_0 = \left(
\dfrac{a_{\rm rh}}{a_0}
\right)^3
B_{\rm rh}^2 L_{\rm rh}  
\end{align} 
where $L_{\rm rh}$ and $L_0$ is the correlation scale at the end of inflation and at the present time, respectively. The above relation must be considered along with a second relation in order to determine the present values of the magnetic intensity and its correlation scale. Reference \cite{Banerjee:2004df} demonstrated that for a large set of initial conditions, the values of $B$ and $L$ at recombination is linked by the relation $B \simeq L H_{\rm rec} \rho^{1/2}$ where $H_{\rm rec}$ is the Hubble parameter at recombination and $\rho$ is the energy density of the fluid particles that couple to the magnetic field. Evolving this relation until today, under the condition that their comoving values stay constant, we find a very general relation~\cite{Banerjee:2004df,Durrer:2013pga}
\begin{align}
\label{B_0_2}
B_0 \simeq  10^{-8} {\rm G} \left(
\dfrac{L_0}{{\rm Mpc}}
\right) \,.
\end{align}
The relations \eqref{B_0_1} and \eqref{B_0_1} are a consequence of the inverse cascade of
the helical field associated to the self-similar evolution. 

One can determine the current values of the magnetic intensity and the
correlation scale by combining Eqs.~\eqref{B_0_1} and \eqref{B_0_1} to obtain
\begin{align}
\label{B0helical}
B_0 &= 10^{-8} {\rm G} \left(
\dfrac{B_{\rm rh}}{10^{-8} {\rm G}}
\right)^{2/3} \left(
\dfrac{L_{\rm rh}}{{\rm Mpc}}
\right)^{1/3}\left(
\dfrac{a_{\rm rh}}{a_0}
\right) \,,
\\
\label{L0helical}
L_0 &= \left(
\dfrac{B_{\rm rh}}{10^{-8} {\rm G}}
\right)^{2/3} \left(
\dfrac{L_{\rm rh}}{{\rm Mpc}}
\right)^{1/3}\left(
\dfrac{a_{\rm rh}}{a_0}
\right) {\rm Mpc} \,.
\end{align} 
Using Eqs.~\eqref{a-scale} and \eqref{T_rh}, the magnetic field intensity and the correlation scales can be obtained in terms of the Hubble rate at the end of inflation and other parameters of the model.


\section{Noise correlations for the helical EM fields}
\label{noise}
In this appendix, we  derive the explicit forms of the quantum noises  arising from the short modes of electromagnetic fields. For the non-helical electromagnetic fields the corresponding results can be found in Refs.~\cite{Talebian:2019opf,Talebian:2020drj,Fujita:2017lfu}. 

Expanding \eqref{sigmaX} in terms of the creation and annihilation operators $a_{\bf k}$ and $a^{\dagger}_{\bf k}$, we find
\begin{equation}
\hat{\sigma}^{_X}_i ({\bf x},t)= -\dfrac{{\dd}k_{\rm c}}{\dd t}\sum_{\lambda}^{}\int \frac{{\rm d}^3k}{(2\pi)^3} \, \delta \big( k-k_{\rm c} \big)  \, e_i^\lambda({\bf k}) \bigg(
X_{k,\lambda}(t) \, a_{{\bf k},\lambda} + X_{k,\lambda}(t)^* \,  a^\dagger_{-{\bf k},\lambda}
\bigg) e^{i{\bf k}.{\bf x}} \, ,
\end{equation}
where $k_{\rm c} \equiv \epsilon \, a(t) H$.
Without loss of generality, we assume $\bfx=r \ \hat{z}$ and consider the wave number $\hat{\boldsymbol k}$ and the polarization vectors $\boldsymbol{e}_\lambda(\hat{\boldsymbol{k}})$ as
\begin{align}
\label{k}
\hat{\boldsymbol k} &=
\Big(
\sin\theta \cos\phi ,\ \sin\theta \sin\phi ,\ \cos\theta
\Big) \,,
\\
\label{e}
\boldsymbol{e}_\lambda(\hat{\boldsymbol{k}}) &= \dfrac{1}{\sqrt{2}}
\Big(
\cos\theta \cos\phi-i\lambda \sin\phi,\ \cos\theta \sin\phi +i\lambda \cos\phi,\ -\sin\theta
\Big) \,
\end{align}
in the Cartesian coordinate. One can easily check that the above satisfies the orthogonality relations \eqref{Orthogonality}. 

To calculate the correlation of the helical noises, we use the fact that for any $\lambda$-dependent function $g_\lambda$, one has
\begin{align}
\sum_{\lambda = \pm}  \ g_\lambda \ e_i^{\lambda}(\hat{\boldsymbol{k}})~e_j^{\lambda *}(\hat{\boldsymbol{k}}) &=
\dfrac{1}{2}\sum_{\lambda = \pm}  \ g_\lambda
\left(
\delta_{ij}-\hat{k}_i \hat{k}_j + i\lambda \ f_{ij}(\theta,\phi)
\right)
\end{align}
where $f_{ij}$ is an anti-symmetric matrix, $f_{ij}=-f_{ji}$, given by
\begin{align}
f_{21}(\theta)=\cos\theta \,,
\hspace{8mm}
f_{32}(\theta,\phi)=\sin\theta \cos\phi\,,
\hspace{8mm}
f_{13}(\theta,\phi)=\sin\theta \sin\phi \,.
\end{align}
Using the following relations
\begin{align}
\int_\Omega {\rm d}\Omega \ e^{ikr\cos\theta}
&=
\int_{\phi=0}^{2\pi}{\rm d}\phi \int_{0}^{\pi} \sin\theta \ {\rm d}\theta \ e^{ikr\cos\theta}= 4\pi \dfrac{\sin(kr)}{kr} \overset{r\rightarrow 0}{=} 4\pi \,,
\\
\int_\Omega {\rm d}\Omega \ e^{ikr\cos\theta} \left(\delta_{ij}-\hat{k}_i\hat{k}_j\right)
&=
\dfrac{8\pi}{3} \ \delta_{ij} \dfrac{\sin(kr)}{kr} \overset{r\rightarrow 0}{=} \dfrac{8\pi}{3} \ \delta_{ij} \,,
\\
\int_\Omega {\rm d}\Omega \ e^{ikr\cos\theta} f_{ij}
&\overset{r\rightarrow 0}{=}0 \,,
\end{align}
one can find that 
\begin{align}
\label{X_correlation}
&&\left \langle
{\hat{\sigma}}^{_X}_i(t_1,\bfx) \ {\hat{\sigma}}^{_X}_j(t_2,\bfx)
\right \rangle
&= \dfrac{1}{18\pi^2} \dfrac{{\rm d}k_{\rm c}^3}{{\rm d}t} \ \sum_{\lambda}^{}\big|X_\lambda(t_1,k_c)\big|^2 \ \delta_{ij} \ \delta(t_1-t_2)  \,.
\end{align}
Using Eq. \eqref{v-mode} as well as the definition of electric and magnetic fields Eq. \eqref{electric}, we find
\begin{align}
\label{electricamplitude}
	\left<{\hat{\sigma}}^{_E}_i(N_1),{\hat{\sigma}}^{_E}_j(N_2)\right> &=\frac{H^2 2^{1-2 n} \varepsilon ^{4-2 n} \Gamma (2 n)^2}{3 \pi ^2 \Mp^2  \abs{\Gamma (n+i \xi )}^2}\cosh(\pi  \xi ) \, \delta_{ij}\delta(N_1-N_2),\quad 0<n<2 \,,
	\\
\label{magneticamplitude}
	\left<{\hat{\sigma}}^{_B}_i(N_1),{\hat{\sigma}}^{_B}_j(N_2)\right> &=\frac{H^2\varepsilon^5(2\varepsilon)^{-2|n-\frac{1}{2}|}\Gamma(2|n-\frac{1}{2}|)^2}{3\pi^2\Mp^2\abs{\Gamma(\frac{1}{2}+|n-\frac{1}{2}|+i\xi)}^2}\cosh(\pi\xi) \, \delta_{ij}\delta(N_1-N_2),\quad 0<n\neq\frac{1}{2}<2 \,,
\end{align}
while for $n=1/2$ we have
\begin{align}
\label{magneticamplitude_n1/2}
\left<{\hat{\sigma}}^{_B}_i(N_1),{\hat{\sigma}}^{_B}_j(N_2)\right> &= \frac{H^2 \varepsilon ^5  \log ^2(\varepsilon )}{3 \pi ^3\Mp^2}\cosh(\pi  \xi ) \, \delta_{ij}\delta(N_1-N_2),\quad n=\frac{1}{2} \,.
\end{align}

\subsection{Disappearance of the quantum nature of the noises}
\label{quantum_nature}

Here, we show that the quantum nature of noises disappear on large scales. To this end we show that the following commutator goes to zero in this limit:
\begin{equation}
	\frac{ [{\hat{\tau}}_{_X},{\hat{\sigma}}_{_X}]}{D_{_X}^2}\rightarrow0
\end{equation}
where ${\hat{\tau}}_{_X}$ is the noise corresponding to the conjugate momentum of $X$ defined by
\begin{equation}
	{\hat{\tau}}_{_X} ({\bf x},t)= -\dfrac{{\dd}k_{\rm c}}{\dd t} \int \frac{\dd^3k}{(2\pi)^3} \, \delta( k-k_{\rm c}) \, e^{i {\bf k}.{\bf x}} \,  \dot{\hat{X}}_k \, .
\end{equation}
If we show that the above relation holds, then it is logical to neglect  the quantum nature of $X$ on large scales, while keeping the effects of the classical noise of the  $X$ field in the analysis.

Using Eq. \eqref{v-mode} as well as the definition of electric and magnetic fields Eq. \eqref{electric}, we obtain the following equations for the  commutator of electromagnetic fields and their conjugate momentum \begin{equation}\label{commuteelectric}
			[{\hat{\sigma}}^{_E}_i(N_1),{\hat{\tau}}^{_E}_j(N_2)]=-\frac{2 i H^2 \xi  \varepsilon ^4 \sinh (2 \pi  \xi )}{3 \pi ^2M_P^2}\delta_{ij}\delta(N_1-N_2),\quad 0<n<2 \, .
		\end{equation}
Comparing \eqref{electricamplitude} and \eqref{commuteelectric} we see that the ratio of the amplitude of the commutator  to the amplitude of noise of  electric field is ${\cal O}(\varepsilon^{2n})$ and can be neglected in the range we are interested. This shows that as far as $\varepsilon \ll 1$ the noises can be treated classically. In the same manner one can write the amplitude of magnetic field and its commutator as follows:
\begin{equation}
[{\hat{\sigma}}^{_B}_i(N_1),{\hat{\tau}}^{_B}_j(N_2)]=-\frac{i H^2 \varepsilon ^5 \cosh (2 \pi  \xi )}{3 \pi ^2 M_p^2}\delta_{ij}\delta(N_1-N_2),\quad 0<n<2 \,.
\end{equation}
We see that the ratio of the commutator to the amplitude of magnetic field is ${\cal O}(\epsilon^{2|n-\frac{1}{2}|})$ when $n\neq \frac{1}{2}$ and ${\cal O}(\log^{-2}(\epsilon))$ when $n=\frac{1}{2}$. Therefore, we conclude that one can neglect the quantum nature of the noises and treat them as classical noises.

With the above property and the disappearance of the quantum nature of the noises,
one can express the quantum noises ${\hat{\sigma}}\textsub{X}(N)$ in terms of the classical normalized white noise $\sigma$ as
\begin{align}
\label{sigmaXX2}
{\hat{\sigma}}^{_X}_i(N) \equiv \DX \,\, \sigma_i(N) \,,
\end{align}
where
\begin{equation}
\langle \sigma_i(N) \rangle = 0 \,,
\hspace{3cm}
\langle \sigma_i(N_1) \, \sigma_j(N_2) \rangle = \delta_{ij} \, \delta(N_1-N_2)\,,
\end{equation}
and for $n\ne 1/2$, the amplitude $\DX$ is given by
\begin{align}
\label{DXX}
D\textsub{X}
&\simeq  \dfrac{\sqrt{2 \cosh(\pi \xi)}}{\pi \sqrt{3} }   \dfrac{\Gamma(2n-1)}{2^{n} \abs{\Gamma(n+i\xi)}} \ \dfrac{H}{\Mp} \varepsilon^{2-n} \times
\begin{cases}
(2n-1) \,, \qquad & X = E
\\
\\
\varepsilon \,,\qquad & X = B
\end{cases}
~\,.
\end{align}
This amplitude approaches to what was obtained in \cite{Talebian:2019opf,Talebian:2020drj} when $\xi \rightarrow 0$ for non-helical electromagnetic noises where the two transverse modes are the same.

\section{Diagonalization}
\label{diagonal}
In this appendix, we solve the coupled Langevin equations
\begin{align}
{\cal B}' &= -(2+n){\cal B}+D\textsub{B} \, \sigma(N) \,,
\\
{\cal E}' &= -(2-n){\cal E}+2 n\gamma {\cal B}+D\textsub{E} \, \sigma(N) \,.
\end{align}
A common way of handling these equations is to look for a change of coordinates or a change of variables which simplifies the problem. We use the diagonal matrix method to solve the equations. Let us write Eqs. \eqref{langvinB2} and \eqref{langvinE2} in the following matrix form
\begin{equation}\label{matrix}
\begin{pmatrix}
{\cal E}'\\ 
{\cal B}'
\end{pmatrix}=C \begin{pmatrix}
{\cal E}\\ 
{\cal B}
\end{pmatrix}+\sigma(N)\begin{pmatrix}
\DE\\
\DB
\end{pmatrix} \, ,
\end{equation}
where $C$ is the matrix of coefficients given by
\begin{equation}
C\equiv  \begin{pmatrix}
n-2& 2n\gamma\\
0& -(2+n)
\end{pmatrix} \, .
\end{equation}
Having the matrix of coefficients at hand, one can easily write the basis transformation matrix as
\begin{equation}
P=\begin{pmatrix}
-\gamma& 1\\ 
1 & 0
\end{pmatrix},
\end{equation}
which is obtained using the eigenvectors of $C$. 

Now according to the fact that any  vector like $V$ in the old basis changes as $\tilde{V}=P^{-1}V$ in the new basis, one can write \eqref{matrix} as
\begin{equation}\label{newbasis}
\begin{pmatrix}
\tilde{{\cal E}}'\\ 
\tilde{{\cal B}}'
\end{pmatrix}=\tilde{C} \begin{pmatrix}
\tilde{{\cal E}}\\ 
\tilde{{\cal B}}
\end{pmatrix}+\sigma(N)\begin{pmatrix}
\DB\\
\DE+\gamma\DB
\end{pmatrix} \, ,
\end{equation}
where \begin{equation}\label{newbasis2}
\tilde{C}\equiv P^{-1}C P=
\begin{pmatrix}
-(2+n)&0\\
0&n-2
\end{pmatrix} \, ,
\end{equation}
and the tilde in the components denote the changed vector. Then in new basis we obtain
\begin{align}
\tilde{{\cal E}}' &= -(2+n)\tilde{{\cal E}}+D\textsub{B} \, \sigma(N) \,,
\\
\tilde{{\cal B}}' &= -(2-n)\tilde{{\cal B}}+(\DE+\gamma\DB) \, \sigma(N) \,.
\end{align}
Note that since the electric and  magnetic fields originate from the same gauge field, the noises $\sigma\textsub{B}$ and $\sigma\textsub{E}$ are not independent, $i.e.$ $\sigma\textsub{B}+\sigma\textsub{E}=(\DB+\DE)\sigma(N)$\footnote{Otherwise, we have $\sigma\textsub{B}+\sigma\textsub{E}=(\DB+\DE)^{1/2} \, \sigma(N)$.}. Then we obtain two decoupled Langevin equations which can be solved easily by the appropriate initial conditions, ${\cal B}(N=0)={\cal B}_0$ and ${\cal E}(N=0)={\cal E}_0$. More specifically, the solutions are given by
\begin{equation}
\begin{pmatrix}
\tilde{\cal E}(N)\\ 
\tilde{\cal B}(N)
\end{pmatrix}=\begin{pmatrix}
{\cal B}_0 \, e^{-(n+2)N}+\DB \,\int_{0}^{N}  e^{(n+2)(N'-N)}\sigma(N') \dd N'\\\\ \\
({\cal E}_0+\gamma {\cal B}_0)e^{(n-2)N}+(\DE+\gamma\DB)\int_{0}^{N} e^{-(n-2)(N'-N)}\sigma(N') \dd N'
\end{pmatrix} \, .
\end{equation}
Now going back to the old basis by $\begin{pmatrix}
{\cal E}\\ 
{\cal B}
\end{pmatrix}=P\begin{pmatrix}
\tilde{{\cal E}}\\ 
\tilde{{\cal B}}
\end{pmatrix}$
we have
\begin{align}
\label{Bsol}
{\cal B}(N) &= {\cal B}_0 e^{-(n+2) N}+\DB \, 
\int_{0}^{N} e^{(n+2) (N'-N)} \ \sigma(N') \, \dd N' \, ,
\\
\label{Esol}
{\cal E}(N) &= e^{(n-2) N} \left(
{\cal E}_0+\gamma {\cal B}_0
\right)
-\gamma  {\cal B}(N)+\left(
\DE +\gamma\DB
\right)
\int_{0}^{N} e^{-(n-2) (N'-N)} \ \sigma(N') \, \dd N' \, .
\end{align}
\section{Probabilistic analysis}
\label{sec:PDF}
In this section, we use another approach to study the Langevin equation \eqref{langvinB2}. Let us recast Eq. \eqref{langvinB2} into the following stochastic differential
equation (SDE)
\begin{eqnarray}
\dfrac{{\rm d}{\cal B}_i(N)}{{\rm d}N} &=& -\mu~{\cal B}_i(N)+ \DB~\xi_i(N)  \,;
\hspace{1cm}
\mu \equiv n+2 \,. 
\label{calB-Langevin}
\end{eqnarray}
We are interested in the regime $\mu>0$ so the above equation describes an Ornstein-Uhlenbeck (OU) process. Therefore the field ${\cal B}_i$ admits a equilibrium state with a long-term mean  and a bounded variance (mean-reverting process) due to the fact that the random force $\DB \, \xi_i$ balances the frictional drift force $-\mu \, {\cal B}_i$~. To be more precise, an OU process is a stationary Gauss-Markov process in which there is the tendency for the system to drift toward the mean value, with a greater attraction when the process is further away from the mean. For this process, the explicit dependence of the mean to the initial conditions is washed out over time and the system can only de described by the drift $\mu$ and the diffusion $\DB$ coefficients. In  other words, the distribution of the random variable $\mathcal{B}_i$ can be described by the normal distribution $\mathbb{N}\Big(0,\dfrac{\DB^2}{2\mu}\Big)$ at $N=N^B_{\rm eq}$. Formally, $N_{\rm eq}^B$ goes to infinity, but we can estimate the equilibrium time as when the relative difference of the field with its equilibrium value drops to a small value say $10^{-2}$.
With this approximation,
we obtain \eqref{NeqB}.

Alternatively, the Fokker-Planck equation associated  with the Langevin equation \eqref{calB-Langevin} can be employed to describe the time evolution of the probability density function (PDF) of ${\cal B}_i (N)$. Consider $f_{{\cal B}_i}(x,N)$ as the 
PDF of the random variable ${\cal B}_i$. Then the associated Fokker-Planck equation is given by
\begin{eqnarray}
\dfrac{\partial f_{{\cal B}_i}(x,N)}{\partial N} = -\mu\dfrac{\partial}{\partial x} \bigg(x f_{{\cal B}_i}(x,N) \bigg) + \dfrac{\DB^2}{2} \dfrac{\partial^2}{\partial x^2}f_{{\cal B}_i}(x,N) \,.
\label{probab_X_i}
\end{eqnarray}
Intuitively, one can think of $f_{{\cal B}_i}(x,N) {\rm d}x$ as the probability of ${\cal B}_i$ falling within the infinitesimal interval $[x,x+{\rm d}x]$. Assuming a stationary probability distribution, $\partial f^{\rm eq}_{{\cal B}_i}/\partial N=0$, the equilibrium solution of Fokker-Planck Eq. \eqref{probab_X_i} is given by
\begin{eqnarray}
\label{f_Xi_station}
f^{\rm eq}_{{\cal B}_i}(x) = \sqrt{\dfrac{\mu}{\pi \DB^2}} ~\mathrm{exp} \Big({-\dfrac{\mu}{\DB^2}}x^2 \Big)\,.
\end{eqnarray}
Using the above 
PDF for the components of the magnetic field ${\cal B}_i$, it is easy to obtain 
PDF of its magnitude ${\cal B}\equiv\left( \sum_{i=1}^{3}{\cal B}_i^2\right)^{1/2}$ as follows:
\begin{eqnarray}
\label{f_X_station}
f^{\rm eq}_{{\cal X}}(x) &=& 4~ \sqrt{\dfrac{\mu^3}{\pi \DB^6}} ~x^2~   \mathrm{exp} \Big({-\dfrac{\mu}{\DB^2}}x^2\Big)\, .
\end{eqnarray}

This density function allows us to calculate the $m$-th moments associated with ${\cal B}$ as follows:
\begin{eqnarray}
\left\langle \mathcal{B}^m  \right\rangle_{\rm eq} &=& \int_{0}^{\infty} {\rm d}x~x^m~f^{\rm eq}_{{\cal B}}(x)=\dfrac{2}{\sqrt{\pi}}\left(
\dfrac{\DB}{\sqrt{\mu}}
\right)^m \, \Gamma\left(
\frac{m+3}{2}
\right)\,,
\label{X_m}
\end{eqnarray}
Moreover, these  
PDFs enable us to calculate the probability of having a given amplitude for the magnetic field in a given range.  The desired range corresponds to the lower and upper bounds on cosmological magnetic fields as given  in Eq. \eqref{B-bound}. Subsequently, 
these bounds are translated  into the interval ${\cal B}_1<{\cal B}<{\cal B}_2$. Therefore, one can calculate the probability of generated magnetic field acquiring a value in the interval determined in \eqref{B-bound},  given by
\begin{align}
\label{Pbminus}
P_{B_{\rm obs}} 
&= \int_{{\cal B}_1}^{{\cal B}_2} {\rm d}x ~ f^{\rm eq}_{\cal B}(x)
\nonumber
\\&=
\mathrm{Erf}\left(y_2\right)-\mathrm{Erf}\left(y_1\right) -\dfrac{2}{\sqrt{\pi}} \left(y_2~ e^{-y_2^2}-y_1~ e^{-y_1^2}\right)
\,,
\end{align}
in which $y_i\equiv \dfrac{\sqrt{\mu}}{\DB}{\cal B}_i$ and $i=1,2$~. In fact, the above is the probability of generating the primordial magnetic field consistent with the observational bound \eqref{B-bound} by the model \eqref{action}, $P_{B_{\rm obs}}(n,\xi,\varepsilon)$. 
The probabilistic interpretation based on the Fokker-Planck equation is a parallel approach to the mechanism of stochastic differential equations presented in Section \ref{BToday}.

It is interesting to obtain the stationary PDF of the electric field. One can write \eqref{Esol2} as follows
\begin{align}
\label{Esol3}
{\cal E}_i(N) + \gamma  {\cal B}_i(N)&= \left(
\DE +\gamma\DB
\right)
\int_{0}^{N} e^{-(n-2) (N'-N)} \ \sigma_i(N') \, \dd W(N') \,.
\end{align} 
As can be seen,
the stochastic variable ${\cal E}_i(N) + \gamma  {\cal B}_i(N)$
satisfies a Gaussian PDF. 
As $\gamma  {\cal B}_i$ is a Gaussian variable then one can easily see that $\left( {\cal E}_i(N) + \gamma  {\cal B}_i(N)\right)-\gamma  {\cal B}_i(N)$ is a Gaussian as well. Therefore by the mean and variance of electric field at the equilibrium state one finds
\begin{eqnarray}
\label{f_Ei_station}
f^{\rm eq}_{{\cal E}_i}(x) = \sqrt{\dfrac{1}{\pi \left<{\cal E}_i^2\right>_{\rm eq}}} ~\mathrm{exp} \Big({-\dfrac{x^2}{\left<{\cal E}_i^2\right>_{\rm eq}}} \Big)
\,,
\end{eqnarray}
where $\left<{\cal E}_i^2\right>_{\rm eq}$ is given by 
\eqref{Eeq}.

\section{ Gravitational waves induced by gauge field and the energy scale of inflation}
\label{GW}
The gauge fields are the additional sources of tensor perturbations, besides the vacuum ones. Which contribution is the dominant  one  depends directly on the model parameter. For $\xi \sim {\cal O}(10)$ where the tachyonic enhancement of the gauge field is significant, then the gravitational wave signal actively sourced by the gauge field is more significant. In this appendix, we study the production of gravitational waves induced by the electromagnetic modes. 

Let us turn on the tensor perturbations of the metric \eqref{metric} via
\begin{align}
\label{hij_metirc}
\dd s^2 = a^2(\tau) \Big[-\dd \tau^2 +  \left(\delta_{ij}+h_{ij}\right) \dd x^i \dd x^j \Big] \,,
\end{align}
in which $h_{ij}(t,{\bfx})$ is the transverse-traceless (TT) tensor perturbation ($\partial_{i}h^{ij}=0 = h^i{}_{i}$). 
The quadratic expansion of the action \eqref{action} for tensor part leads to~\cite{Barnaby:2011qe,Barnaby:2011vw}
\begin{align}
S^{(2)}_{\rm t} = \dfrac{\Mp^2}{8}\int \dd^3x \ \dd \tau \ a^2 \bigg[
h'_{ij}{}^2-(\partial_k h_{ij})^2-\dfrac{4a^2}{\Mp^2}h_{ij} S_{ij}
\bigg] \,;
\hspace{1cm}
S_{ij} = E_iE_j + B_i B_j \,.
\end{align}
Therefore the equation of motion for the tensor modes is given by
\begin{equation}\label{hEoM}
h_{ij}''+2\mathcal{H}h_{ij}'-\nabla^2h_{ij}
=-\dfrac{2a^2}{\Mp^2}S_{ij}\,,
\end{equation} 
where $\mathcal{H}=a'/a$ is the comoving Hubble parameter. The equation of motion \eqref{hEoM} is solved by separating $h_{ij}$ into a vacuum modes $h_{ij}^{(0)}$, the solution of the homogeneous equation, and a sourced mode $h_{ij}^{(s)}$. The modes produced by the
gauge quanta are statistically uncorrelated with those from the vacuum.

In the absence of source, the power spectrum has the standard form
\begin{align}
{\cal P}_h^{(0)} = 
\dfrac{2H^2}{\pi^2 \Mp^2} \,.
\end{align}
Since we are interested in super-horizon solutions for the sourced modes, we simply neglect the negative helicity mode and the gradient term in the Fourier expansion of \eqref{hEoM}. Then the tensor mode
is given by
\begin{align}
h_+ \simeq \dfrac{-2 S_{\rm eq}}{\Mp^2 H^2} {\cal N} \,;
\hspace{2cm}
S_{\rm eq} \simeq \Mp^2 H^2 \left<{\cal E}_i^2\right>_{\rm eq} \, ,
\end{align}
in which ${\cal N} \sim 60$ is $e$-folding number of inflation and $\left<{\cal E}_i^2\right>_{\rm eq}$ is defined in \eqref{Eeq}. Note that we work in the parameter space $1/2<n<2$ and have also neglected the contribution of magnetic field due to the suppression factor $\varepsilon^2$ in \eqref{DXX}  in comparison to the electric part. Therefore we find that
\begin{align}
{\cal P}_h^{(s)} \simeq 4 {\cal N}^2 \left<{\cal E}_i^2\right>_{\rm eq}^2  \,.
\end{align}
The two contributions add up  in the power spectrum, yielding 
\begin{align}
{\cal P}_h = {\cal P}_h^{(0)} + {\cal P}_h^{(s)} \,.
\end{align}
Due to the production of the gauge quanta, 
the tensor-to-scalar ratio $r_{\rm t} \equiv {{\cal P}_h}/{\cal P}_\zeta$ can be estimated as
\begin{align}
\label{rt}
r_{\rm t} \simeq \dfrac{{\cal P}_h^{(0)}}{{\cal P}_\zeta} + \dfrac{4 {\cal N}^2 \left<{\cal E}_i^2\right>_{\rm eq}^2}{{\cal P}_\zeta} \,.
\end{align}

From CMB observations  \cite{Akrami:2018odb}, the scalar power spectrum is given by ${\cal P}_\zeta \simeq 2.1 \times 10^{-9}$, while 
the constraint on the tensor-to-scalar ratio $r_{\rm t}$ is $r_{\rm t} < 10^{-2}$. The above relation along with \eqref{Eeq} for $1/2<n<2$ beside the observational values for $r_{\rm t}$ and ${\cal P}_\zeta$, lead to a relation for the Hubble parameter during inflation which we denote by the dimensionless parameter $h$,
\begin{align}
\label{H_Mp}
\dfrac{H}{\Mp} &\equiv h(\xi,n,r_{\rm t}) \,.
\end{align}
For large enough $\xi$, we obtain $h \propto e^{-\pi \xi}\left({\cal P}_\zeta r_{\rm t}\right)^{1/4}$ which is  consistent with \eqref{H_Caprini}. Having obtained the 
Hubble parameter during inflation, one can estimate the energy scale of inflation defined as $\rho_{\rm inf}^{1/4} \equiv \left(3\Mp^2H^2\right)^{1/4}$. In Fig.~\ref{fig:rho_inf}, we have plotted the value of inflationary energy scale as a function of  $\xi$. We see that for $\xi > 4$ the energy scale of inflation decreases rapidly.



\small
\bibliography{references}
\bibliographystyle{JHEPNoTitle} 

\end{document}